\documentclass{article}

\PassOptionsToPackage{numbers, compress}{natbib}

 \usepackage[preprint]{neurips_2026}

\usepackage[utf8]{inputenc} 
\usepackage[T1]{fontenc}    
\usepackage{hyperref}       
\usepackage{url}            
\usepackage{booktabs}       
\usepackage{graphicx}       
\usepackage{amsmath}        
\usepackage{amsfonts}       
\usepackage{nicefrac}       
\usepackage{microtype}      
\usepackage{xcolor}         
\usepackage{xspace}         
\usepackage{caption}        
\usepackage{wrapfig}        
\usepackage{placeins}       
\usepackage[most]{tcolorbox}
\tcbset{claimstyle/.style={
  colback=gray!12,
  colframe=gray!40,
  arc=4pt,
  boxrule=0.6pt,
  left=0pt, right=0pt, top=0pt, bottom=0pt,
  width=0.99\linewidth,
  before upper={\setlength{\parskip}{1pt}},
}}

\title{CCL-Bench 1.0: A Trace-Based Benchmark for LLM Infrastructure}

\author{%
  Eric Ding\thanks{\texttt{ericding@cs.cornell.edu}} \quad
  Byungsoo Oh \quad
  Bhaskar Kataria \quad
  Kaiwen Guo \quad
  Jelena Gvero \quad \\
  \textbf{Abhishek Vijaya Kumar} \quad
  \textbf{Arjun Devraj} \quad
  \textbf{Lindsey Bowen} \quad
  \textbf{Atharv Sonwane} \quad \\
  \textbf{Emaad Manzoor} \quad
  \textbf{Rachee Singh} \\
  \\ 
  Cornell University
}

\usepackage{tikz}
\usetikzlibrary{decorations.pathreplacing}
\newcommand{\sysname}{CCL-Bench\xspace}
\newcommand{\syssearch}{CCL-Search\xspace}
\providecommand{\vs}{vs. }
\providecommand{\ie}{\emph{i.e.,} }
\providecommand{\eg}{\emph{e.g.,} }

\providecommand{\alltoall}{\textsc{AllToAll}\xspace}
\providecommand{\allreduce}{\textsc{AllReduce}\xspace}
\providecommand{\allgather}{\textsc{AllGather}\xspace}

\providecommand{\reducescatter}{\textsc{ReduceScatter}\xspace}

\providecommand{\workload}{\textit{workload}\xspace}

\providecommand{\fullinfra}{\textit{full infrastructure}\xspace}
\newenvironment{packeditemize}{\begin{list}{$\bullet$}{\setlength{\itemsep}{0.1pt}\addtolength{\labelwidth}{0pt}\setlength{\leftmargin}{\labelwidth}\setlength{\listparindent}{\parindent}\setlength{\parsep}{1pt}\setlength{\topsep}{0pt}}}{\end{list}}
\providecommand{\myparab}[1]{\vspace{0pt}\noindent\textbf{#1} }

\begin{document}

\maketitle

\vspace{-2em}
\begin{abstract}
  Evaluative claims about LLM infrastructure ---``workload X is fastest on hardware Y with software Z''---depend on a complex configuration space spanning hardware accelerators, interconnect bandwidth, software frameworks, parallelism plans, and communication libraries. Current infrastructure evaluation benchmarks publish a small set of end-to-end numbers that do not explain why one configuration outperforms another. 
We present \sysname, a trace-based benchmark that addresses the limitations of existing benchmarks by recording reusable evidence for every ML workload. Each contributed data point in \sysname packages an execution trace, a YAML workload card, and the launch scripts. We have developed a community-extensible toolkit to compute fine-grained compute, memory, and communication efficiency metrics from this evidence. 
Using \sysname, we surface three claims that summary-statistic benchmarks cannot support: (i) higher compute-communication overlap can coincide with longer training step time and reveal inefficient parallelization choices, (ii) doubling TPU interconnect bandwidth yields a much higher end-to-end improvement in step time than doubling GPU interconnect bandwidth on small and medium workloads, and (iii) the best-tuned configuration on one training framework can run up to 3$\times$ slower than the best-tuned configuration on a peer framework on identical hardware.

\end{abstract}
\vspace{-1em}

\section{Introduction}
\label{sec:introduction}

Several broad audiences rely on LLM infrastructure benchmarks to make critical decisions. Hardware vendors who design accelerators (\eg GPUs, TPUs~\cite{googlecloudtpu}, Trainium~\cite{fu2024distributed}, and MTIA~\cite{meta_mtia_2023})
need to ensure that representative workloads perform well on their hardware. Software framework and library developers who build training engines, serving engines, communication libraries, and compilers that run on accelerators (\eg Megatron-LM~\cite{shoeybi2019megatron}, vLLM~\cite{kwon2023efficient}, NCCL~\cite{nccl_docs}, XLA~\cite{sabne2020xla}) need to localize where their software spends time on a given hardware platform. Production operators who deploy LLM training jobs or serving endpoints (\eg AWS~\cite{aws}, Azure~\cite{microsoftazure}) need to find configurations that meet latency or throughput budgets while minimizing costs.

\myparab{Limitations of Current Benchmarks.}
Current benchmarks are insufficient for three reasons.
First, \emph{evaluation metrics are frozen at experiment time.} Benchmarks like MLPerf~\cite{reddi2020mlperf,mattson2020mlperf}, LLM-Perf~\cite{hf_optimum_llm_perf_leaderboard}, and other vendor leaderboards publish a small set of end-to-end summary numbers. To understand the same workload along a new evaluation axis, like communication breakdown~\cite{hta2024}, compute utilization~\cite{li2020xsp}, straggler severity~\cite{stragglar}, or memory-transfer overhead~\cite{ren2021zerooffload}, requires another full experiment. Second, current benchmarks provide \emph{limited insights for performance enhancement.} While current benchmarks support performance comparisons of existing software-hardware stacks~\cite{semianalysis2026inferencex,chung2025ml}, they provide limited insights on where improvements should be made to close a given performance gap. Third, \emph{tuning effort is invisible.} A reported win on public leaderboards often reflects a well-tuned engine against a stock baseline, and the ranking can change once both sides are tuned. 

\myparab{Outcome \vs Explanation.}
The common thread across these limitations is that current benchmarks report only \emph{outcomes} but not \emph{explanations}. An outcome tells the reader which infrastructure combination is fastest on a given workload. An explanation tells the reader which component contributed to the win, whether the ranking would survive a different metric or configuration, and where the next bottleneck lies. Without an execution record rich enough to support explanation, every new evaluation metric requires a new experiment, often on hardware the reader does not own.

\begin{figure}[t]
  \centering
  \includegraphics[width=0.95\linewidth, height=0.15\textheight]{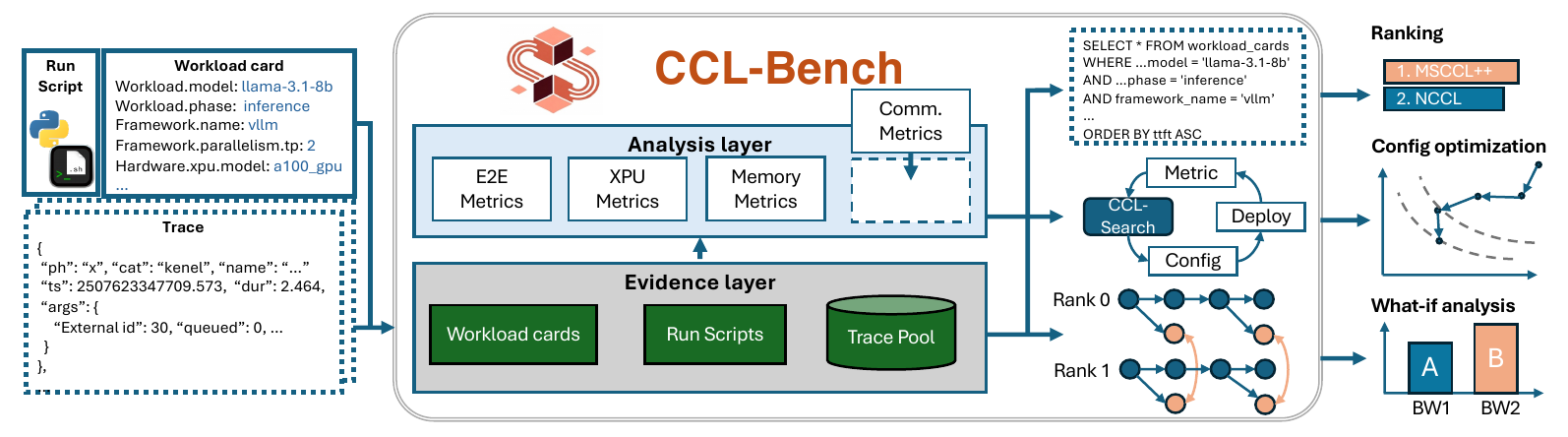}
  \caption{\small{\sysname overview: a standardized trace plus workload card recording each run, a metric toolkit computing fine-grained metrics, and downstream analysis and optimization plug-ins.}}
  \label{fig:overview}
  \vspace{-1.2em}
\end{figure}

We present \sysname, a trace-based benchmark for LLM infrastructure that records fine-grained execution evidence sufficient to explain observed performance. \sysname has two components.

\begin{packeditemize}
    \item \myparab{An evidence-based schema.} Every \sysname submission includes three artifacts: an execution trace~\cite{kineto2026} that records operators, kernels, communication events, timestamps, and per-rank activity; a YAML workload card that specifies the \workload{} (model family, batch
    size, sequence length, step count) and names the \fullinfra under test (hardware architecture, training or serving engine, parallelism configuration, collective library tuning, compiler); and the run scripts that launched the experiment. Together, these artifacts let a reader attribute measured performance to specific infrastructure components without rerunning the workload.

    \item \myparab{A community-extensible metric toolkit.}
    \sysname includes an open-source library of analysis tools that are portable across XPU types. A tool is any function that takes a trace and a workload card and returns a scalar metric. The current library covers MFU, compute utilization, compute-communication overlap, memory-transfer overhead, end-to-end step time, and a novel hardware-resource utility metric. New tools can be contributed and applied retroactively.
\end{packeditemize}

\myparab{Evaluations enabled by \sysname.}
\sysname supports two evaluation axes. \emph{Software Infrastructure} comparisons fix the \workload{} and the hardware architecture, vary one software system component, and match the rest of the execution plan 
(\eg NCCL~\cite{nvidia2025nccltuning} \vs MSCCL++~\cite{microsoft2026mscclpp} on the same GPU cluster running Llama-3.1-8B). 
\emph{Hardware Infrastructure} comparisons fix the \workload{} and the XPU budget and let each platform run its architecture-native deployment (\eg PyTorch with NCCL on GPU \vs MaxText~\cite{maxtext} with XLA on TPU). \sysname's protocol of trace collection lets a reader attribute an observed performance gap to compute, memory, or collectives. Beyond these comparisons, \sysname supports three use cases that current benchmarks do not enable.

\begin{packeditemize}
  \item \emph{Post-hoc metric extension.}
        A new evaluation metric can apply retroactively to every compatible trace in the \sysname pool. To demonstrate this, we add a communication traffic-volume metric to explain a performance gap observed between two benchmark entries.

  \item \emph{Trace-driven what-if analysis.}
        Empirical traces in \sysname can feed popular distributed ML simulators~\cite{won2023astrasim,wang2024simai} to estimate how iteration time changes under different interconnect bandwidth, topology, or collective algorithm assumptions. Grounding the simulation in measured kernel timings avoids the calibration gap of purely simulation-based performance models.

       \item \emph{Automated configuration optimization.}
        \sysname integrates CCL-Search, an LLM-agent-based optimizer that automates configuration tuning on any hardware--software infrastructure. Given a workload, a target infrastructure, and an optimization objective (\eg minimize step time, or balance latency against accelerator cost), CCL-Search iteratively proposes configurations, executes them on hardware, collects traces, and evaluates the objective using the metric toolkit to refine its next proposal. The resulting configuration policy and every intermediate trial are recorded as \sysname benchmark entries, so the tuning effort is verifiable and reproducible.

\end{packeditemize}

\myparab{Technical claims.}
Using \sysname, we test technical hypotheses and make the following claims:
\begin{packeditemize}
  \item Higher compute-communication overlap does not always reduce LLM training step time, because the parallelism choices that increase overlap can also increase collective traffic volume.
  \item Doubling TPU interconnect bandwidth yields up to $100\times$ higher end-to-end step-time improvement than doubling GPU interconnect bandwidth on small and medium workloads, while large GPU training workloads benefit more from a larger scale-up domain.
  \item A training framework with its best-found parallelism configuration can run up to $3\times$ slower on a peer framework on identical hardware with the same workload.
\end{packeditemize}

We make \sysname's workload cards and metric toolkits publicly available at \url{https://github.com/cornell-sysphotonics/ccl-bench}. \sysname is hosted at \url{https://cclbench.ai/}.

\section{Benchmarking Infrastructure for Large Language Models}

\begin{wrapfigure}{r}{0.45\textwidth}
\centering
\vspace{-1em}
\begin{tikzpicture}[
  layer/.style={
    draw=black, thick, rounded corners=2pt,
    minimum width=5cm, minimum height=0.5cm,
    font=\footnotesize\sffamily, align=center
  },
  lbl/.style={
    font=\scriptsize\sffamily, text=gray!70!black, anchor=east
  },
  brace/.style={
    decorate, decoration={brace, amplitude=5pt}, thick
  }
]
  \node[layer, fill=blue!8]   (hw) at (0,0)    {\textbf{Hardware Platform} \\
    {\scriptsize Accelerator, memory, interconnect, topology}};
  \node[layer, fill=orange!8] (sw) at (0,0.9)  {\textbf{Software} \\
    {\scriptsize Compiler, communication library, framework}};
  \node[layer, fill=green!8]  (wl) at (0,1.8)  {\textbf{Workload} \\
    {\scriptsize Model family, phase, batch size, sequence length}};

  \draw[brace] ([xshift=-4pt]hw.south west) -- ([xshift=-4pt]sw.north west)
    node[midway, sloped, above=6pt, font=\scriptsize\sffamily]
    {\textbf{Infrastructure}};

\end{tikzpicture}

\caption{\small{LLM infrastructure evaluated by \sysname. The workload defines
what computation must happen. The infrastructure---hardware platform and
software---determines how it is executed.}}
\label{fig:infra-layers}
\vspace{-1em}
\end{wrapfigure}
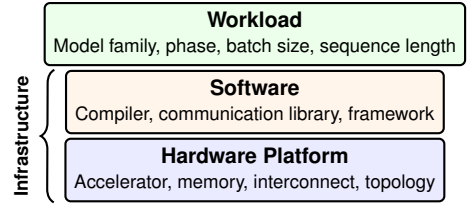

\sysname evaluates the infrastructure that executes LLM workloads. This section defines how the software, hardware and LLM workload compose (Figure~\ref{fig:infra-layers}).

\myparab{Workload }specifies the computation task to be performed, \ie the model family and size, quantization, the phase (training or inference), the batch size, and the sequence length. \sysname targets open-weight models that can be hosted by a third party for evaluation on their infrastructure.
The workload determines the arithmetic computation of an experimental run.

\myparab{Software.}
The software maps a workload onto a hardware platform. It comprises, from lowest to highest level of (1) compiler support for graph capture, operator fusion, and kernel generation (\eg TorchInductor~\cite{pytorch2} and Triton~\cite{tillet2019triton}), (2) the communication library (\eg NCCL, MSCCL++, XLA collectives~\cite{openxla_operation_semantics}) and (3) the training or serving framework together with its parallelism plan (\eg TorchTitan~\cite{liang2024torchtitan}, Megatron-LM, vLLM, SGLang, or MaxText). The framework is configured with Data Parallelism (DP), Tensor Parallelism (TP), Pipeline Parallelism (PP), Expert Parallelism (EP) degrees, micro-batch size, pipeline schedule, and memory-management policies like activation recomputation, ZeRO stage, offload, and cudagraph~\cite{narayanan2021megatronlm,ren2021zerooffload,rajbhandari2020zero,rajbhandari2021zero,pytorch_cuda_graphs}.

\myparab{Hardware platform} is the physical environment on which the workload executes. This includes the accelerator type and model (GPU, TPU, or other XPU), the memory hierarchy, the server-scale interconnect (\eg NVLink~\cite{nvidianvlink}, UALink~\cite{ualinkspec}), and the datacenter-level network topology (\eg Torus~\cite{jouppi2023tpuv4}, Dragonfly~\cite{kim2008technology}, Rail-optimized~\cite{qian2024alibaba}).

\myparab{Infrastructure.}
Infrastructure is the combination of hardware platform and software. It is the object of study in \sysname: every evaluative claim names a workload and a piece of infrastructure. 
\subsection{Existing Benchmarks}

Machine learning benchmarks such as LLM-Perf and MLPerf~\cite{reddi2020mlperf,mattson2020mlperf,semianalysis2026inferencex, chung2025ml, jiang2024moe,artificialanalysis_llm_leaderboard,hf_optimum_llm_perf_leaderboard,coleman2017dawnbench,zhang2019ai,zhu2018benchmarking,adolf2016fathom} run full models and report scalar summaries that are used for performance rankings.
They are useful for understanding current infrastructure's performance via specific metric suites (\eg query latency in InferenceX~\cite{semianalysis2026inferencex}, energy consumption in ML.Energy~\cite{chung2025ml}).
Parameterized studies~\cite{wang2020systematic} broaden the metric or workload surface but remain coupled to the metrics selected before execution.

Infrastructure providers have published component-level benchmark artifacts with new releases. Hardware vendors report product-specific results with limited metrics~\cite{googlecloud2026tpuv5p,googlecloud2025aihypercomputerinference, nvidia2025blackwellmlperf,nvidia2025blackwellultramlperf, nvidia2025trainingmlperf,amd2025mlperfinference,amd2025mlperftraining, nvidia2023h100mlperf}. Communication libraries focus on collective microbenchmarks and tuning studies~\cite{nvidia2026nccltests,nvidia2025nccltuning,hwang2026mscclpp, microsoft2026mscclpp}. Serving engines including vLLM and SGLang release throughput and latency benchmarks scoped to individual releases~\cite{vllm2024perfupdate,vllm2026benchmarks,sglang2026benchserving, sglang2026docs}. These artifacts expose useful LLM-specific metrics (TTFT, TPOT, collective bandwidth) but compare against selected baselines on a single platform. Their logging, metric calculation, and tuning records are not standardized across contributors. Automated tuning strategies for parallelism and communication~\cite{zheng2022alpa,miao2022galvatron,jia2019flexflow} address part of the reproducibility gap but require significant re-engineering to port across software engines or hardware platforms.

\myparab{Trace-based systems.}
One family of trace-based systems treats the trace as the workload itself, providing compact, replayable objects for co-design~\cite{chakra2026}, serving request streams~\cite{wang2025burstgpt}, or scheduling and provisioning studies~\cite{stojkovic2025dynamollm,patel2024splitwise}. A second family uses traces as evidence for analyzing the original execution. Holistic Trace Analysis (HTA)~\cite{hta2024} exemplifies this approach, computing kernel breakdowns, communication-computation overlap, and trace diffs over Kineto~\cite{kineto2026} traces. \sysname follows this second direction and extends it into a benchmarking framework with standardized workloads, comparison axes, and a shared trace pool for cross-infrastructure metric analysis.

\myparab{Simulators and performance models.}
Simulators for model architecture, network topology, and hardware~\cite{won2023astrasim,wang2024simai,bansal2022jahs}, along with tools for training/inference configuration planning and deployment~\cite{agrawal2024vidur,cho2024llmservingsim,zhang2024llmcompass, bang2024vtrain,zheng2022alpa}, enable design-space exploration without running every configuration on physical hardware. Their outputs complement but do not substitute for evidence grounded in observed executions, and they depend on empirical calibration that trace-based benchmarks can provide.

\subsection{Limitations of Existing Benchmarks}

Existing LLM infrastructure benchmarks face three challenges that limit their diagnostic value and leave them functioning only as scoreboards.

\myparab{Benchmark artifacts are metric-specific.}
Suites like MLPerf~\cite{reddi2020mlperf}, LLM-Perf~\cite{hf_optimum_llm_perf_leaderboard}, and vendor leaderboards answer only the metrics chosen by the benchmark owner, but not metrics introduced later by a library developer, hardware designer, or operator. 
These suites are scoreboards that define what is measured, but they do not expose a reusable artifact from which new metrics or bottleneck attributions can be derived post-hoc.
Follow-up questions about communication breakdown~\cite{hta2024}, compute utilization~\cite{li2020xsp}, or memory-transfer overhead~\cite{ren2021zerooffload} require rerunning the workload and laborious tuning because the published result does not contain the evidence needed to compute them.
Cross-referencing different benchmarks is also not possible due to the large configuration space.

\myparab{Rankings do not identify the bottleneck.}
End-to-end benchmark scores are useful for comparisons, but they provide little guidance for improving one stack or investing in the next hardware revision. After observing a performance gap, a user needs to know whether the limiting factor is collective communication, memory bandwidth, placement of parallelism domains, or saturation of a particular network fabric. Scalar benchmark reports discard the per-rank timeline~\cite{mattson2020mlperf,reddi2020mlperf,semianalysis2026inferencex} and configuration context needed to make that attribution.

\myparab{Configuration search is outside the benchmark record.}
LLM performance is a result of environment-specific choices like TP, DP, PP, micro-batch size, collective-library settings, compiler options, and memory policies~\cite{zheng2022alpa,nvidia2025nccltuning}. Existing benchmarks often report the final configuration, but not the search path, failed trials, or objective used to choose it. This makes a published win difficult to reproduce and makes it hard to judge whether two systems were compared at similarly optimized operating points. Requiring every participant to perform this tuning manually also raises the cost of contributing new benchmark entries.

\section{\sysname overview}
\label{sec:overview}

\begin{wraptable}{r}{0.51\linewidth}
  \centering
  \scriptsize
  \vspace{-6em}
  \setlength{\tabcolsep}{3pt}
  \caption{\small{Workload-card fields (subset).}}
  \label{tab:terminology}
  \begin{tabular}{p{0.16\linewidth}p{0.51\linewidth}p{0.20\linewidth}}
    \toprule
    Role & Template field & Example \\
    \midrule
    Workload       & \shortstack[l]{\texttt{workload.model.}\\\texttt{model\_family}} & \texttt{llama-3.1-8b} \\
    Workload       & \texttt{workload.model.phase} & \texttt{training} \\
    Workload       & \texttt{workload.data.batch\_size} & \texttt{4} \\
    Workload       & \texttt{workload.data.seq\_len} & \texttt{8192} \\
    Architecture   & \shortstack[l]{\texttt{workload.hardware.}\\\texttt{xpu\_spec.model}} & \texttt{nvidia\_a100} \\
    Architecture   & \shortstack[l]{\texttt{workload.hardware.}\\\texttt{network\_topo.topology}} & \texttt{slingshot} \\
    System         & \shortstack[l]{\texttt{Model-executor.}\\\texttt{framework.name}} & \texttt{torchtitan} \\
    System         & \shortstack[l]{\texttt{Model-executor.}\\\texttt{model\_plan\_parallelization}} & \shortstack[l]{\texttt{DP shard=2,}\\\texttt{TP=4, PP=2}} \\
    System         & \shortstack[l]{\texttt{Model-executor.}\\\texttt{communication\_library.name}} & \texttt{NCCL} \\
    System         & \shortstack[l]{\texttt{Model-executor.}\\\texttt{protocol\_selection}} & \texttt{rocev2, p2p} \\
    \bottomrule
  \end{tabular}
  \vspace{-3em}
\end{wraptable}

\sysname (Figure~\ref{fig:overview}) records evidence, instead of final evaluation metrics, from LLM workloads. Each experiment contributes to \sysname an execution trace, a workload card describing the workload and infrastructure under test, and the scripts that launched it. Anyone can later compute new metrics on the same evidence without rerunning the experiment, and submit those metrics as separately versioned tools to our open-source repository.

\subsection{What \sysname records}
\label{sec:overview:data}


\myparab{Trace.} \sysname collects LLM execution traces in JSON format, including Kineto traces~\cite{kineto2026} and JAX/XLA profiler traces collected via XProf~\cite{xprof_jax_profiling2026}. Kineto, the tracing backend used by the PyTorch Profiler, supports NVIDIA and AMD GPUs, Intel XPU, and HPU backends~\cite{kineto2026}, while the JAX Profiler/XProf provides trace collection for JAX workloads on CPU, GPU, and TPU through XLA runtime events~\cite{xprof_jax_profiling2026}. For each iteration on every rank, these traces record the operators the model executes, the kernels and communication events each operator launches across one or more streams, and the timestamp of every event, supporting both end-to-end performance metrics and fine-grained trace-level breakdowns. Auxiliary traces, like Nsight Systems traces~\cite{nvidiansightsystems}, further provide hardware-specific counters not exposed by Kineto. We provide details about the trace format and collection overhead in Appendix~\ref{app:trace}.

\myparab{Workload card.} The YAML card records the workload, the architecture, and the software stack so that the trace becomes interpretable evidence. Provenance fields (\texttt{version}, \texttt{description}, \texttt{hf\_url}, \texttt{trace\_url}, \texttt{contributor}) identify the experiment and its author. Workload fields record phase, MoE (Mixture-of-expert model~\cite{fedus2022switch}) flag, model family, floating point precision, iteration count recorded, parameter and layer counts, batch size, sequence length, and dataset. Architecture fields record XPU model, device counts, driver version, and network topology and bandwidth. System fields record the framework, compiler, parallelism plan (DP/TP/PP/CP/EP and pipeline microbatch), and communication library and protocol. Metric source fields list the trace types and any auxiliary artifacts the metric tools need. Table~\ref{tab:terminology} shows representative fields; the full schema is in Appendix~\ref{app:schema}.

\myparab{Run scripts} are required in a submission for reproducibility. We discuss how contributors collect traces to capture accurate infrastructure performance for different workload types in Appendix~\ref{app:collection_method}.


\subsection{How \sysname analyzes evidence}
\label{sec:overview:analysis}

A tool $i$ in \sysname is a function from a workload card and a trace to a scalar metric $f_i : \mathcal{W} \times \mathcal{T} \;\longrightarrow\; \mathbb{R}$, where $\mathcal{W}$ is the space of workload cards and $\mathcal{T}$ is the space of execution traces.
A suite of $n$ tools applied to the same evidence yields a performance profile $
  \mathbf{m}(w, \tau) \;=\; \bigl(f_1(w,\tau),\; f_2(w,\tau),\; \ldots,\; f_n(w,\tau)\bigr) \;\in\; \mathbb{R}^n$, 
where $(w, \tau) \in \mathcal{W} \times \mathcal{T}$ is a single submission.
No single scalar captures infrastructure behavior; the vector $\mathbf{m}$ jointly characterizes compute efficiency, memory throughput, and communication overhead.

\myparab{Toolkit.} The shipping toolkit covers four categories. \emph{Efficiency} reports average step time and Model FLOPs utilization (MFU)~\cite{chowdhery2023palm}. \emph{Compute} reports compute unit coverage and primary kernel timespan. \emph{Memory} reports host-device bandwidth and memory-transfer overhead. \emph{Communication} reports collective bandwidth, communication fraction (amount of time spent in communication), and compute-communication overlap. The toolkit also reports MoE fraction for mixture-of-experts models and TTFT/TPOT for inference. Appendix~\ref{app:tools} catalogs the full set with details for MFU, memory-transfer overhead, and compute-communication overlap.

\subsection{How users interact and contribute}
\label{sec:overview:submit}


\myparab{Evidence interface.}
A contributor runs a workload from \sysname's rolling workload set on the infrastructure under test, collects traces, and uploads them with a workload card and run scripts via the web UI\@.
Rule-based checks verify metadata correctness---minimum iteration count, complete environment reporting, and schema validity.
\sysname then applies the current tool suite and publishes the entry once the submission and metric results pass review.

\myparab{Tool interface.}
A contributor submits a tool along with its category, supported trace types, and expected output. Maintainers check compatibility before admission.
Once admitted, \sysname applies the tool to every compatible trace already in the pool, so a new metric immediately covers all existing workloads without rerunning any experiment.
If a tool is found to be incorrect, \sysname retracts the version, flags all affected entries, and recomputes those metrics against the corrected tool.
Contributors can also challenge an existing entry with a counter-trace or a new tool. The dispute is resolved on the public submission thread (\eg a GitHub pull request).

\myparab{Result access.}
The leaderboard is a two-dimensional table---submissions as rows, metrics as columns. Entries can be clustered by workload---with selectable rows for cross-infrastructure comparison. A website screenshot is attached (Figure~\ref{fig:ui-overview} in Appendix~\ref{app:ui}).
Workload cards, run scripts, and the tool suite are publicly available in the \sysname GitHub repository.
Access to raw traces is gated on contributing a trace for the same workload, incentivizing pool growth. 

\section{Case studies and insights from \sysname}
\label{sec:cases}

We use \sysname to evaluate LLM infrastructure across two hardware
platforms and multiple software configurations. The first is the NERSC Perlmutter supercomputer~\cite{nerscperlmutterarchitecture}, where each node contains four A100 GPUs connected by NVLink~3.0 at 300\,GB/s unidirectional bandwidth (scale-up domain), and nodes connect through a Slingshot-11 fabric at 200\,Gbps (scale-out domain). The second is Google TPU~v6e~\cite{googletpuv6e}, where eight chips form a node and 32 nodes form a pod (scale-up domain) in a 2D torus topology with 100\,GB/s unidirectional inter-chip interconnect (ICI) bandwidth. Software infrastructure spans communication libraries (NCCL, MSCCL++, XLA collectives) and training and serving engines (Megatron-LM, TorchTitan, MaxText, vLLM, SGLang). Table~\ref{tab:leaderboard-workloads} in Appendix~\ref{app:workloads} lists the workload suite we use to put the infrastructure combinations under test. We show how \sysname enables making technical claims that cannot be derived from coarser LLM infrastructure benchmarks. Cross-system and cross-architecture comparisons can be found in Appendix~\ref{app:results}.

\subsection{Perspective of a framework and library developer}

\begin{figure}[t]
  \centering
  \begin{minipage}[t]{0.56\linewidth}
    \centering
    \includegraphics[width=\linewidth]{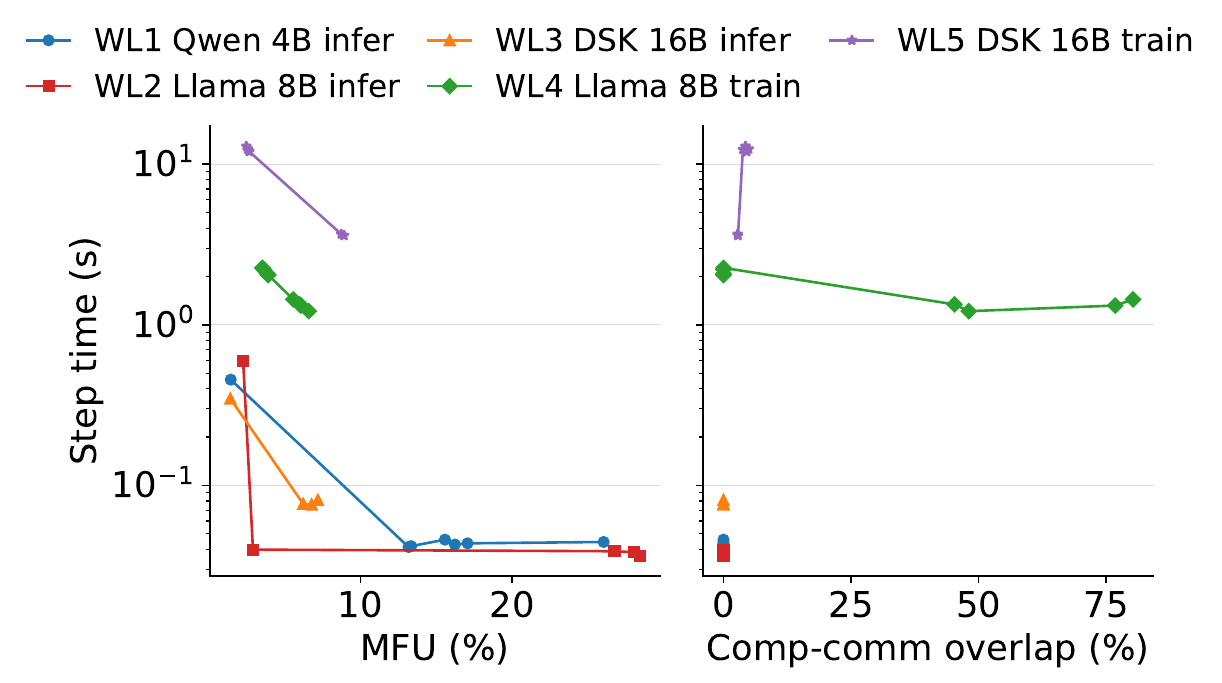}
    {\small\textbf{(a)}}
  \end{minipage}\hfill
  \begin{minipage}[t]{0.40\linewidth}
    \centering
    \includegraphics[width=\linewidth]{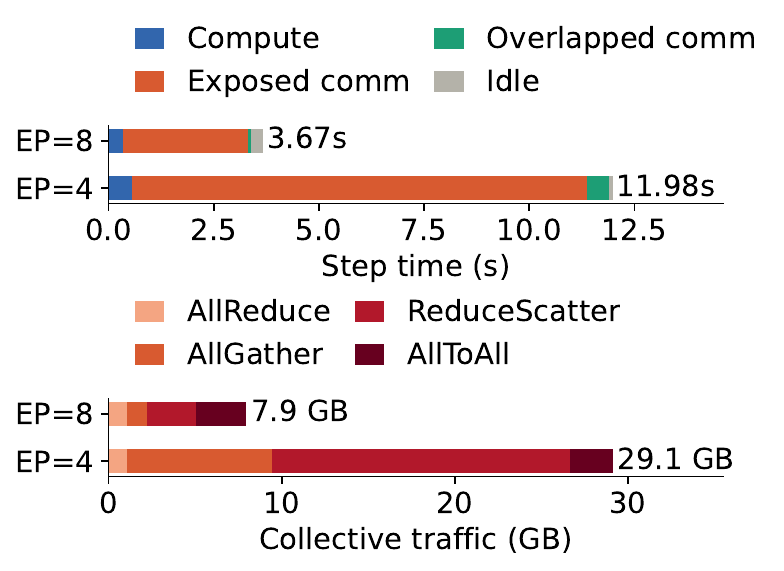}
    {\small\textbf{(b)}}
  \end{minipage}
  \vspace{-0.8em}
  \caption{\small{(a) MFU vs.\ step time and compute-comm overlap vs.\ step time
    on A100 GPU (Perlmutter). Each point is one \sysname entry.
    (b) Step-time and communication traffic-volume break-down for
    WL5 DeepSeek-V3-16B EP=4 and EP=8 MoE training runs (TP=4, DP=2, PP=1).}
  \label{fig:cross-system-mfu-step}}
  \vspace{-1.6em}
\end{figure}

A framework developer tuning a distributed LLM workload typically optimizes for two metrics:  
MFU (model FLOPs utilization) and compute-communication overlap.
Both metrics are commonly used as proxies for lower step time under a fixed
workload and hardware configuration: higher MFU indicates that more of the
hardware's peak compute is being converted into model FLOPs, while higher
compute-communication overlap reduces exposed communication on the critical
path~\cite{chowdhery2023palm,hashemi2018tictac}.

\sysname's trace pool lets us test whether these assumptions hold across configurations.

Figure~\ref{fig:cross-system-mfu-step}(a, left) confirms the first
assumption. Across training and inference workloads in \sysname, higher MFU generally yields lower step time. Inference workloads on vLLM and SGLang reach higher MFU than training workloads on TorchTitan, which require further configuration tuning to achieve comparable utilization. The second assumption, however, does not hold. Figure~\ref{fig:cross-system-mfu-step}(a, right) shows that for DeepSeek-V3-16B MoE training (WL5), higher compute-communication overlap coincides with \emph{worse} step time. A traditional summary-statistic benchmark would report only the step time, but \sysname's execution traces help us diagnose the reason for this counterintuitive finding. 

We develop new metric tools for the diagnosis. First, we attribute the step time to compute \vs \emph{exposed communication} \vs \emph{overlapped communication}. Second, we capture the traffic volume of different collectives in each step. We apply these tools to the existing traces without rerunning any experiment. We find that experiments with a higher degree of compute--communication overlap had used a small expert-parallelism (EP) degree relative to the total GPU count (\eg EP\,=\,4 on 8 GPUs), and experiments with lower overlap had used a larger EP degree (\eg EP\,=\,8). While compute--communication breakdown (Figure~\ref{fig:cross-system-mfu-step}(b)) shows that communication dominates the step time in both configurations, the traffic-volume breakdown shows that workloads with smaller EP domains generate significantly more \reducescatter and \allgather traffic. This increase arises because a smaller EP domain replicates experts across the data-parallelism domain, requiring fully-sharded data parallelism to gather weights and synchronize gradients. Although these collectives can be partly overlapped with computation, the additional communication volume outweighs the overlap benefit and increases overall step time.

This finding can motivate developers to explore expert and data parallelism co-optimization opportunities, balancing expert \alltoall communication with data parallelism collectives. The workload trace collection  and post-hoc metric extension in \sysname enables this finding.

\begin{center}
  \begin{tcolorbox}[claimstyle]
    \textbf{Claim 1:} Higher computation-communication overlap does not always imply lower LLM training step latency; it may instead reveal inefficient parallelization choices.
  \end{tcolorbox}
\end{center}

\subsection{Perspective of a hardware vendor}

A hardware designer evaluating whether to invest in faster interconnects between accelerators needs to estimate how much end-to-end performance a bandwidth upgrade would buy for a given workload. Traditionally, computer architects use simulators to answer such questions~\cite{chakra2026,won2023astrasim,wang2024simai}. However, these simulators rely entirely on analytical models and lack grounding in real execution behavior. \sysname bridges this gap by combining empirical traces with simulation frameworks.

\begin{wrapfigure}{r}{0.6\linewidth}
  \centering
  \includegraphics[width=\linewidth]{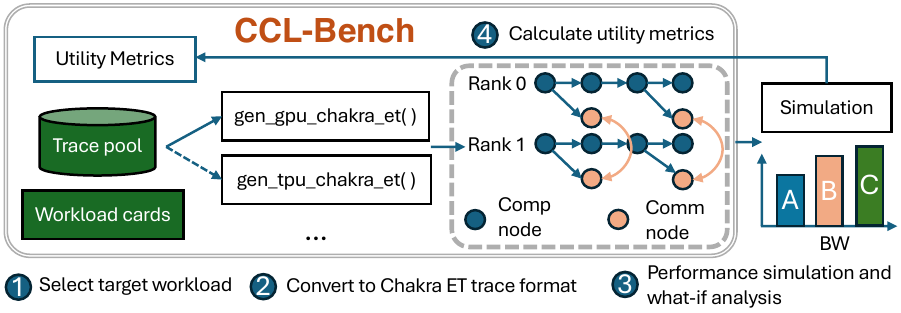}
  \caption{\small \sysname trace-driven what-if pipeline: empirical traces feed a network simulator that estimates step time and the utility of doubling a hardware resource.}
  \label{fig:ccl-utility}
  \vspace{-1em}
\end{wrapfigure}

\myparab{From traces to simulation.} \sysname traces record per-kernel timing, collective types, group IDs, and buffer sizes for every rank. We convert each trace into a Chakra~\cite{chakra2026} execution graph that captures the dependencies between computation and communication across ranks. Compute kernels become \texttt{COMP\_NODE}s with measured durations; collectives become \texttt{COMM\_COLL\_NODE}s parameterized by type, message size, and communicator group; and idle intervals become gap nodes that preserve the original timeline. Edges enforce CUDA/XLA stream order and rank-local sequencing (Figure~\ref{fig:ccl-utility}).

We feed these execution graphs to Astra-Sim~\cite{won2023astrasim}, a distributed ML workload simulator that models network topologies with configurable bandwidth, latency, and collective algorithms. Using Astra-Sim, we replay the computation nodes at their empirical durations and re-simulate the communication nodes under a target network configuration, producing a new end-to-end step time estimate. This pipeline preserves the measured compute behavior of the original run while letting the network parameters vary. We build conversion pipelines for both GPU (Kineto) and TPU (XLA) traces. We use this pipeline to study how improvements in interconnect bandwidth impact iteration time for workloads run on GPUs \vs TPUs. Appendix~\ref{app:sim} includes additional results.


\myparab{Utility metric.}
For a hardware resource $r$ (\eg scale-out network bandwidth, scale-up network bandwidth, or the size of scale-up domain---where all XPUs are interconnected by high-bandwidth scale-up network), let $T$ be the baseline step time and $T_{2\times}(r)$ be the simulated step time after doubling $r$ while holding the workload and all other infrastructure parameters fixed (Figure~\ref{fig:ccl-utility}). We define the \emph{utility} of resource $r$ as the resulting fractional step-time improvement:
\setlength{\abovedisplayskip}{2pt}
\setlength{\belowdisplayskip}{2pt}
\begin{equation}
  \mathrm{Utility}(r) = \frac{T - T_{2\times}(r)}{T} \times 100\%.
  \label{eq:utility}
\end{equation}
A small value indicates that the workload gets little benefit from upgrading $r$, while a large value indicates a bottleneck where the doubling buys meaningful end-to-end speedup. The metric answers an operator-facing upgrade question. Given the workloads already running on the current cluster, which infrastructure upgrade is most worth the spend?

\begin{figure}[t]
  \centering
  \includegraphics[width=\linewidth]{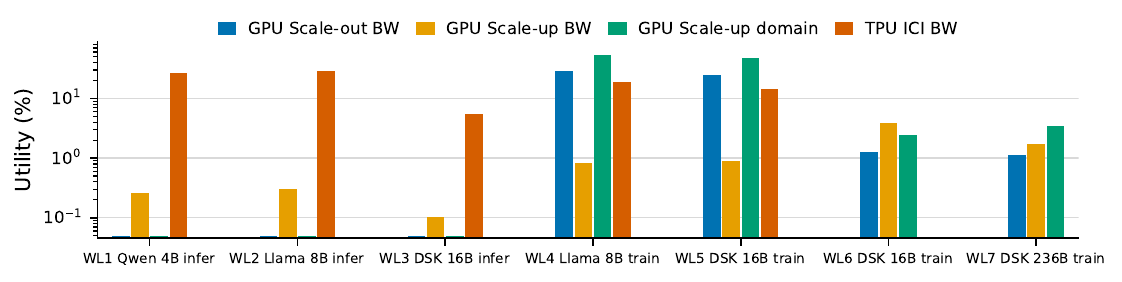}
  \vspace{-2em}
  \caption{\small{Utility metrics for scale-out bandwidth, scale-up/ICI bandwidth, and scale-up domain size on Perlmutter A100 Slingshot and TPUv6e Torus. We are not able to run WL6,7 on TPUs due to resource limits.}}
  \label{fig:utility-metrics}
  \vspace{-1.5em}
\end{figure}

\myparab{GPU \vs TPU bandwidth utility.}
Figure~\ref{fig:utility-metrics} reports utility for representative workloads on Perlmutter A100 (GPU, three axes: scale-out bandwidth, scale-up bandwidth, scale-up domain size) and TPUv6e (TPU, ICI bandwidth). In GPU-based cluster architectures, the best upgrade is workload-dependent. Doubling scale-out bandwidth helps communication-heavy multi-node training (up to 28.7\%) but does not improve single-node inference. Doubling the scale-up domain size provides the largest gains for WL4 and WL5 (53.9\% and 46.7\%) by shifting collective traffic from the slower scale-out fabric onto the faster scale-up fabric. Larger training workloads (WL6 at 128 GPUs, WL7 at 256 GPUs) show low utility (3.9\% and 3.4\%) because their step time is compute-bound. TPU workloads, by contrast, draw high utility from ICI bandwidth doubling for both training and inference, up to $102.57\times$ the matched GPU scale-up bandwidth utility for inference and up to $22.82\times$ for training, due to TPU's lower baseline bandwidth (100GBps ICI vs. 300GBps NVLink) and lower network connectivity (torus vs. fully-connected topology). The contrast indicates that TPU ICI is a more useful investment per bandwidth doubling on small and medium workloads, while GPU clusters benefit more from scale-up domain expansion at model-training scale.

\begin{center}
  \begin{tcolorbox}[claimstyle]
    \textbf{Claim 2:} Doubling TPU ICI bandwidth yields higher end-to-end utility than doubling GPU scale-up bandwidth for small workloads (up to $100\times$ better for inference and $22\times$ for training). For medium-to-large GPU training workloads, investing in a larger scale-up domain is the more useful upgrade than improving the bandwidth.
  \end{tcolorbox}
\end{center}

\subsection{Perspective of a production operator}

A production operator deploying an LLM training job must choose parallelism degrees, micro-batch sizes, activation checkpointing, \emph{etc.}, before launching a run. This deployment configuration space is large, interdependent, and framework-specific (\ie a setting that performs well on one training engine may perform poorly on another). In practice, operators rely on heuristic guidelines and manual iteration, and the tuning effort behind a published benchmark result is rarely recorded or reproducible.

We build an agentic configuration optimizer, \syssearch, that leverages \sysname's evidence and analysis layers (\eg traces, metrics, and prior runs) to progressively discover high-quality configurations through guided experimentation, while recording the entire search process as reproducible benchmark artifacts.

\begin{wrapfigure}{r}{0.65\linewidth}
  \centering
  \vspace{-1em}
  \includegraphics[width=\linewidth]{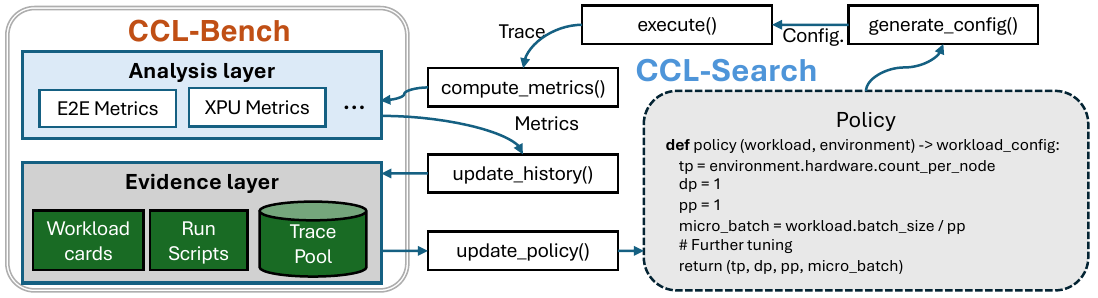}
  \caption{\small \syssearch loop. \texttt{update\_policy} is an LLM
    call. Other steps are local/tool execution.}
  \label{fig:ccl-loop}
  \vspace{-1em}
\end{wrapfigure}

\myparab{How \syssearch works.}
\syssearch is an LLM-agent-based optimizer that iteratively explores the configuration space for a distributed LLM workload (Figure~\ref{fig:ccl-loop}). The user specifies an optimization objective as one or a combination of \sysname metrics (\eg training step time, inference throughput, or a composite objective) and defines the configuration knobs the agent may vary. Like the ADRS framework~\cite{cheng2025barbarians}, \syssearch maintains a policy program, \texttt{generate\_config}, that maps workload and infrastructure context to a concrete configuration (\eg DP, TP, PP, EP degrees, micro-batch size, activation checkpointing). The policy is represented as a programmatic template that is iteratively rewritten by the LLM. Each iteration proceeds in four steps: (1)~\texttt{generate\_config} proposes a configuration, (2)~the configuration is executed on the target hardware, (3)~\sysname collects a trace and computes the objective score via \texttt{compute\_metric}, and (4)~\texttt{update\_policy} invokes an LLM with the score and the full execution history to revise \texttt{generate\_config} for the next round. 

Unlike black-box optimizers, \syssearch enables the LLM to reason over structured configuration semantics (\eg TP/DP tradeoffs, pipeline depth) and trace-level execution feedback, allowing it to avoid invalid, resource-constrained, or inefficient configurations and adapt its strategy based on observed execution behavior. Moreover, every explored configuration produces a complete \sysname benchmark entry---workload card, trace, and score---so the entire search history is preserved for analysis, comparison, and public submission.

\begin{figure*}[t]
  \centering
  \begin{minipage}[t]{0.63\linewidth}
    \centering
    \includegraphics[width=\linewidth]{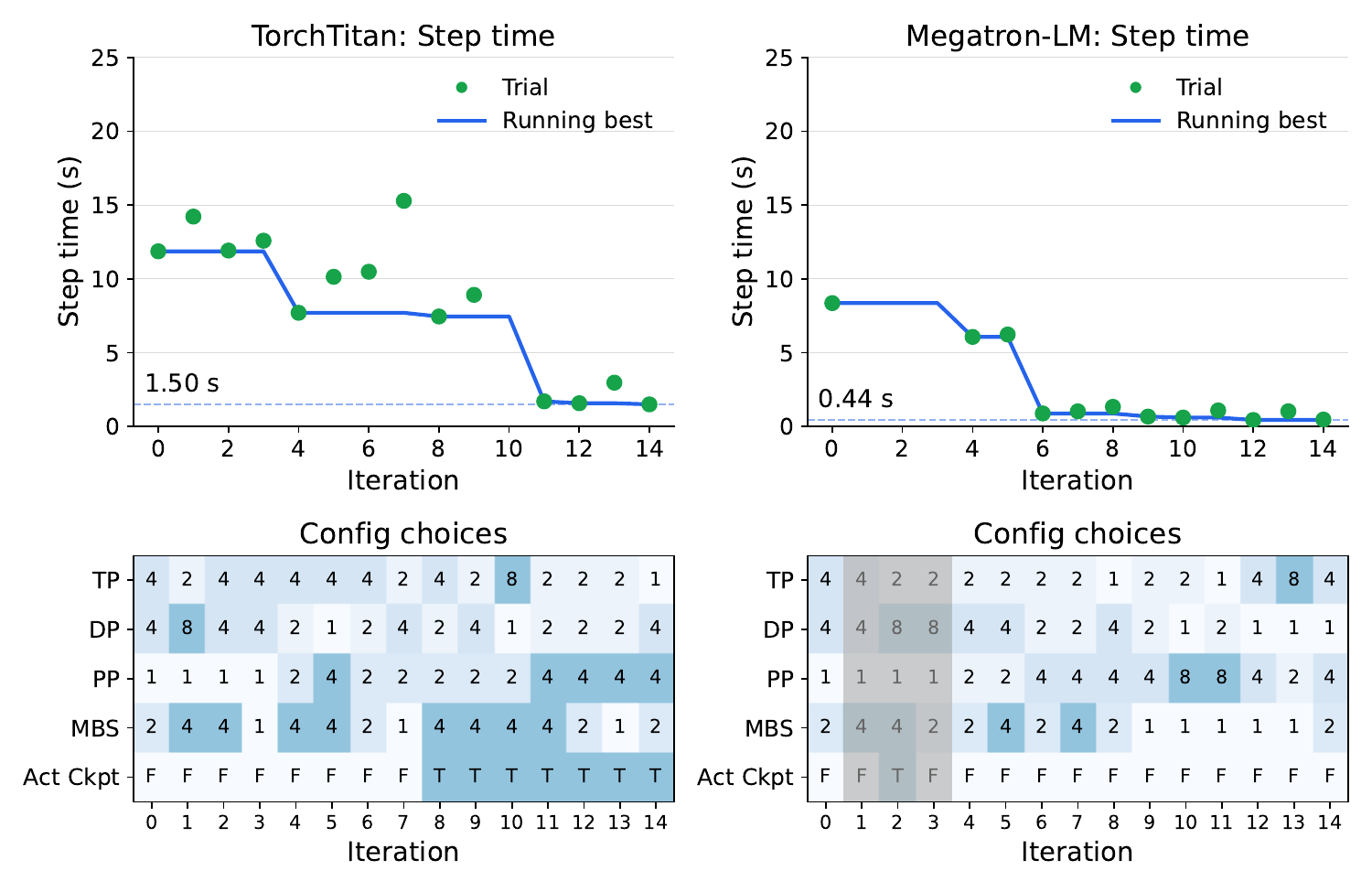}
    \textbf{(a)}
  \end{minipage}
  \hfill
  \begin{minipage}[t]{0.34\linewidth}
    \centering
    \includegraphics[width=\linewidth]{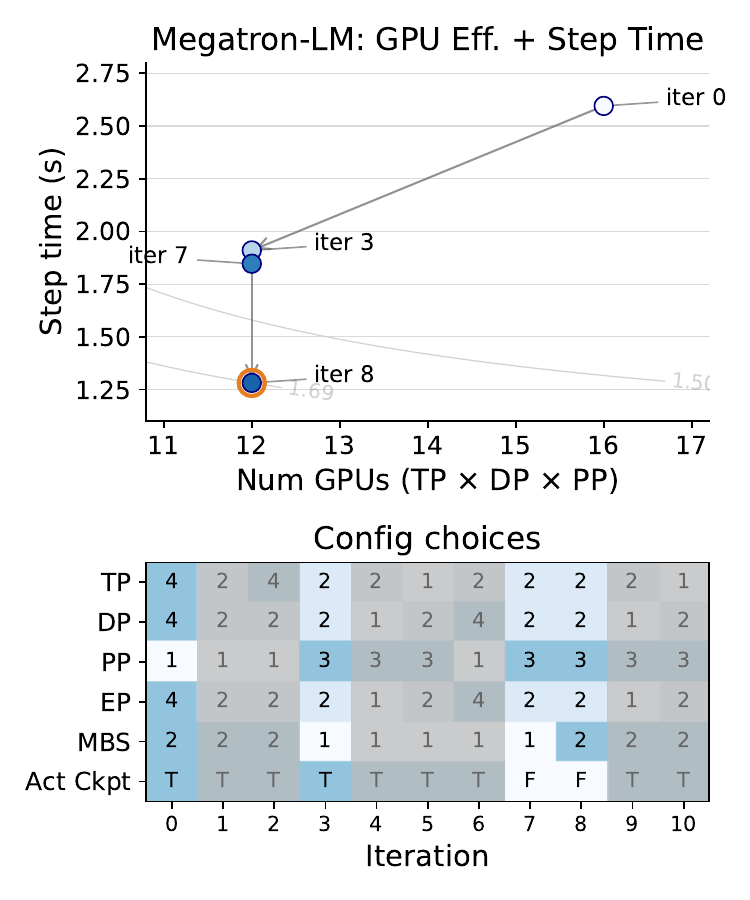}
    \textbf{(b)}
  \end{minipage}
  \caption{\small \syssearch results. Bottom panels show explored
    TP, DP, PP, micro-batch size, and activation-checkpointing choices.
    Grayed-out boxes indicate failed runs. (a)~Step-time objective:
    \syssearch reduces step time by up to $8\times$ on TorchTitan and
    $19\times$ on Megatron-LM within 15 iterations using $Score =
    -avg\_step\_time$. (b)~Composite objective: \syssearch finds the
    Pareto-frontier using $Score = w \times {T_0}/{T} + (1 -
    w) \times {N_0}/{N}$ on Megatron-LM. $T$ is average step time, $N =$ num. XPUs, and $w=0.5$.}
  \label{fig:agent-optimizer-step-time}
  \label{fig:agent-optimizer-pareto}
  \vspace{-1em}
\end{figure*}

\myparab{Framework-specific tuning optima.}
Figure~\ref{fig:agent-optimizer-step-time}(a) shows \syssearch benchmarking two training engines, TorchTitan and Megatron-LM, on Llama-3.1-8B (WL4, 16 Perlmutter GPUs) over 15 iterations across five knobs (TP, DP, PP, micro-batch size, activation checkpointing). The key finding is that each framework's optimum lies at a different point in configuration space. TorchTitan reaches its best step time of 1.50\,s at TP\,=\,1, DP\,=\,4, PP\,=\,4, while Megatron-LM reaches 0.44\,s at TP\,=\,4, DP\,=\,1, PP\,=\,4---a $3.4\times$ gap. Applying TorchTitan's best configuration to Megatron-LM yields 1.3\,s, which is 15\% faster than TorchTitan but still $3\times$ slower than Megatron-LM's own optimum. Thus, configuration policies do not transfer across frameworks. An operator who tunes on one engine and assumes the same settings will work on another leaves significant performance on the table.

\myparab{Composite objectives.}
Figure~\ref{fig:agent-optimizer-pareto}(b) shows \syssearch optimizing for a composite objective, $Score = w \times T_0/T + (1-w) \times
N_0/N$, which balances step time ($T$) against accelerator count ($N$) relative to a baseline ($T_0, N_0$), with $w=0.5$. This formulation rewards configurations that simultaneously reduce latency and hardware cost. On DeepSeek-V3-16B (WL5), the agent identifies Pareto-optimal configurations that trade off latency against hardware cost, enabling cost-aware infrastructure comparisons. Additional runs in Appendix~\ref{app:optimizer} confirm CCL-Search's stability across repeated trials. Since \syssearch records every trial as a \sysname benchmark entry, the tuning process becomes reproducible. An operator on a different cluster can rerun the same policy, inspect the resulting traces, and verify whether the published optimum transfers to their environment.

\begin{center}
  \begin{tcolorbox}[claimstyle]
    \textbf{Claim 3:} Deploying the same workload on different training
    engines requires framework-specific tuning. A configuration that is
    optimal for one framework can be up to $3\times$ slower than the
    best-found configuration on another.
  \end{tcolorbox}
\end{center}

\section{Discussion and Limitations}
\label{sec:cases:discussion}

\myparab{Coverage.}
\sysname 1.0 covers small-to-medium open-source models on GPU and TPU for training and batch inference. Future versions will add larger models, additional accelerators (\eg Trainium), online serving with request-level latency, and accuracy-affecting optimizations (quantization, sparsity, speculative decoding). Exhaustive coverage of the workload--infrastructure cross product is impossible by construction, so \sysname adopts a rolling, theme-driven submission model (\eg communication backends, interconnect fabrics, MoE collectives) on top of a workload- and infrastructure-agnostic trace protocol that lets the suite extend without schema changes.

\myparab{Tools and plug-ins.}
The same trace artifact supports post-hoc metric extension and downstream simulation-based what-if analysis. Building new trace-facing simulator plug-ins, or wiring \sysname into existing performance models, is a natural next step~\cite{liang2025lumos}.

\myparab{Trace storage at scale.}
A single multi-rank trace can reach 10\,GB, and 500$+$ accelerator runs approach 100\,GB. Scaling the leaderboard will require a submission-selection policy that prioritizes diversity across models, platforms, and parallelism, together with compressed or sampled trace formats that preserve the fields the metric tools consume. Additional discussion, including handling of proprietary information in traces, is in Appendix~\ref{app:impact}.


\section{Conclusion}
\label{sec:conclusion}
\sysname is a first step toward trace-based LLM infrastructure evaluation.
Originating from a university class project,
\sysname has collected over 100 workloads, spanning over 7 model architectures, 7 
frameworks, and 3 hardware environments.
The benchmarking platform features a community-extensible metric toolkit, a trace-replay pathway to hardware simulators for what-if analysis, and CCL-Search, a benchmarking protocol for automating
infrastructure configuration optimization.
We provide case studies that demonstrate \sysname surfaces findings that coarser, headline-number benchmarks cannot.
We invite the community to contribute matched-workload traces and analysis tools to broaden cross-system and cross-architecture coverage.

{
\small
\bibliographystyle{plainnat}
\bibliography{references}
}


\appendix


\section{Trace collection}
\label{app:trace}

\sysname accepts two families of trace format, JSON traces and NSYS traces, with JSON traces as the
preferred source for general metrics.
\textbf{Chrome-trace JSON files}
(\texttt{traceEvents} array) are produced on both GPU (Kineto/CUDA backend)
and TPU (XLA backend).  We prioritize this format because it is portable
across accelerator families, exposes the operator, kernel, communication,
rank, stream, and timestamp structure needed by most \sysname metrics.

GPU and TPU Chrome-trace JSON files share the same container format but differ in
several aspects of their event schema:
\begin{itemize} 
\item{Timestamps.}
GPU events carry \texttt{ts}/\texttt{dur} wall-clock timestamps in microseconds
(CUDA host clock); TPU events use \texttt{device\_offset\_ps} /
\texttt{device\_duration\_ps} in picoseconds (XLA device clock), requiring unit
conversion before any cross-platform time comparison.

\item{FLOPs reporting.}
XLA records \texttt{model\_flops} directly in trace-event arguments, so the MFU
tool can sum device-reported FLOPs; GPU traces carry no equivalent field, so
FLOPs must be estimated from the workload-card model architecture.
\item{Step boundaries.}
GPU traces delimit iterations with \texttt{ProfilerStep\#N} spans emitted by the
PyTorch profiler; TPU traces use \texttt{\$core.py:331~step} (or equivalent XLA
step markers) as iteration boundaries.

\item{Collective naming.}
GPU collectives appear as NCCL kernel names (\eg \texttt{ncclAllReduce\_\ldots});
TPU collectives appear as XLA HLO category strings (\eg \texttt{all-reduce},
\texttt{all-gather}) attached to device-side events.

\item{Memory-transfer events.}
GPU traces expose explicit \texttt{memcpy}/\texttt{memset} CUDA kernel records in
separate streams; TPU traces encode DMA as paired \texttt{copy-start.$k$} /
\texttt{copy-done.$k$} async events identified by device-side timestamps.
\item{Stream model.}
GPU traces interleave events across multiple CUDA streams (compute,
communication, memory copy); TPU traces typically present a single device
execution timeline per chip, with host annotations on a separate thread.
\end{itemize}
\textbf{NSYS} traces are NVIDIA Nsight Systems SQLite databases exported
from \texttt{nsys profile}.  They provide NVIDIA-specific CUDA, CUPTI, and
interconnect details that are useful for hardware-counter metrics, but
they are less general across accelerators and are treated as auxiliary
evidence when a metric needs information not present in the Torch trace.

JSON trace can be collected with low profiling overhead in routine benchmark
runs.
To show this, we train a 1.88B test model using tensor parallelism on one machine with 
Intel Xeon Gold 6438M2 CPU and 2 NVIDIA L40 GPUs, and find that 
the tracing overhead is only 0.22\% (averaged over 6 steps), with a coefficient of 
variation of 2.67\%.

\section{Workload-card and trace schema}
\label{app:schema}

\begin{table}[!htbp]
  \centering
  \scriptsize
  \setlength{\tabcolsep}{2.5pt}
  \caption{Workload-card schema fields from the YAML template.}
  \label{tab:full_workload_card}
  \begin{tabular}{p{0.15\linewidth}p{0.58\linewidth}p{0.20\linewidth}}
    \toprule
    Role & Template field & Example \\
    \midrule
    Provenance & \texttt{version}; \texttt{description}; \texttt{hf\_url}; \texttt{trace\_url}; \texttt{contributor}; \texttt{contact} & \shortstack[l]{\texttt{1}; \texttt{https://}\\\texttt{huggingface.co/...}} \\
    Workload model & \texttt{workload.model.phase}; \texttt{moe}; \texttt{granularity}; \texttt{model\_family}; \texttt{precision}; \texttt{epochs}; \texttt{iteration} & \shortstack[l]{\texttt{training}; \texttt{false};\\\texttt{model\_fwd\_bwd\_pass};\\\texttt{llama-3.1-8b}; \texttt{bf16}} \\
    Model architecture & \texttt{workload.model.model\_arch.num\_params}; \texttt{num\_params\_embedding}; \texttt{num\_layers}; \texttt{num\_heads}; \texttt{head\_dim} & \shortstack[l]{\texttt{16380544000};\\\texttt{28}; \texttt{16}; \texttt{128}} \\
    Workload data & \texttt{workload.data.batch\_size}; \texttt{seq\_len}; \texttt{input\_len}; \texttt{output\_len}; \texttt{dataset} & \shortstack[l]{\texttt{4}; \texttt{8192};\\\texttt{1024}; \texttt{128}; \texttt{c4}} \\
    Hardware network & \texttt{workload.hardware.network\_topo.topology}; \texttt{bandwidth\_gbps} & \shortstack[l]{\texttt{slingshot};\\\texttt{[200, 2000]}} \\
    Hardware XPU & \texttt{workload.hardware.xpu\_spec.type}; \texttt{model}; \texttt{total\_count}; \texttt{count\_per\_node}; \texttt{workload.hardware.driver\_version} & \shortstack[l]{\texttt{GPU}; \texttt{nvidia\_a100};\\\texttt{16}; \texttt{4}; \texttt{cuda\_12.4}} \\
    Executor framework & \texttt{Model-executor.framework.name}; \texttt{compiler\_tool\_selection} & \shortstack[l]{\texttt{torchtitan};\\\texttt{plain\_pytorch}} \\
    Parallelization & \texttt{Model-executor.model\_plan\_parallelization.dp\_replicate}; \texttt{dp\_shard}; \texttt{tp}; \texttt{pp}; \texttt{cp}; \texttt{ep}; \texttt{pp\_mb} & \shortstack[l]{\texttt{1}; \texttt{2}; \texttt{4};\\\texttt{2}; \texttt{1}; \texttt{1}; \texttt{1}} \\
    Communication & \texttt{Model-executor.communication\_library.name}; \texttt{version}; \texttt{env}; \texttt{Model-executor.protocol\_selection} & \shortstack[l]{\texttt{NCCL}; \texttt{2.14.3};\\\texttt{NCCL\_IB\_QPS...};\\\texttt{[rocev2, p2p]}} \\
    Metric source & \texttt{metric\_source.traces}; \texttt{metrics\_specific\_trace} & \shortstack[l]{\texttt{[nsys, json]};\\\texttt{[memory\_trace]}} \\
    \bottomrule
  \end{tabular}
\end{table}

\FloatBarrier

\section{Run scripts and trace collection method.}
\label{app:collection_method}
Contributors attach the launch scripts and environment definitions, and run at least five training iterations (steps) per workload in steady state (after 5 initial warm-up iterations that allow CUDA kernels to compile and memory to settle~\cite{pytorch_cuda_graphs}). For inference, contributors first send a small number of warm-up batches so that the serving engine completes CUDA graph capture and KV-cache initialization~\cite{vllm_bench_latency}, then collect traces that cover both the prefill pass (prompt processing) and at least 128 steady-state decode steps (autoregressive token generation). The iteration count enters the workload card so downstream analyses can quantify the variance behind a published number. Contributors are not required to perform full workload training or fine-tuning, as \sysname does not collect accuracy metrics.

\section{Metric tool catalog}
\label{app:tools}

Each \sysname metric is implemented as a self-contained Python tool
under \texttt{tools/<metric>/}.  The dispatcher (\texttt{tools/main.py})
reads the workload card's \texttt{metric\_source.traces} field and
invokes the appropriate backend for each trace format.
Table~\ref{tab:metric-catalog} lists all currently supported metrics.

\begin{table}[!htbp]
\centering
\caption{\sysname 1.0 metric catalog. ``Dir'' indicates whether a higher ($\uparrow$)
or lower ($\downarrow$) value is preferred.
\textbf{P} = PyTorch/Kineto JSON (GPU) or XLA profiler JSON (TPU);
\textbf{P}$_\text{gpu}$ = GPU-only Kineto; \textbf{N} = Nsight Systems SQLite.}
\label{tab:metric-catalog}
\footnotesize
\setlength{\tabcolsep}{3.5pt}
\begin{tabular}{llcclp{5.4cm}}
\toprule
\textbf{Metric key} & \textbf{Label} & \textbf{Unit} & \textbf{Dir} & \textbf{Src} & \textbf{Description} \\
\midrule
\multicolumn{6}{l}{\textit{Model execution}} \\
\texttt{avg\_step\_time} & Step Time & s & $\downarrow$ & P
  & Average wall-clock time per engine iteration (full fwd+bwd for training; decode step for inference). \\[2pt]
\texttt{mfu} & MFU & \% & $\uparrow$ & P
  & Model FLOPs Utilization: observed FLOP/s $\div$ (peak FLOP/s $\times$ device count). \\[2pt]
\texttt{ttft} & TTFT & s & $\downarrow$ & P
  & Time to first token: latency of the prefill pass. Inference only. \\[2pt]
\texttt{tpot} & TPOT & s & $\downarrow$ & P
  & Time per output token: mean decode-step latency after the prefill. Inference only. \\
\midrule
\multicolumn{6}{l}{\textit{Compute}} \\
\texttt{mean\_sm\_coverage} & SM Coverage & \% & $\uparrow$ & P$_\text{gpu}$, N
  & Average SM occupancy across all GPU kernels, weighted by kernel duration. \\[2pt]
\texttt{dominant\_kernel\_concentration} & Top Kernel & \% & $\downarrow$ & P, N
  & Fraction of total GPU/TPU time in the single most time-consuming kernel or HLO category. \\[2pt]
\texttt{compute\_bound\_fraction} & Compute-Bound & \% & $\uparrow$ & P, N
  & Fraction of GPU time in compute-bound kernels (SM coverage $>$70\% and duration $>$10\,\textmu s). \\[2pt]
\texttt{memory\_bound\_fraction} & Memory-Bound & \% & $\downarrow$ & P, N
  & Fraction of GPU/TPU time in memory-bound kernels (SM coverage $<$50\%). \\[2pt]
\texttt{moe\_fraction} & MoE Fraction & \% & $\downarrow$ & P$_\text{gpu}$, N
  & Fraction of GPU time in MoE-specific kernels (expert compute, routing, dispatch/combine). \\
\midrule
\multicolumn{6}{l}{\textit{Memory}} \\
\texttt{average\_memory\_bandwidth} & Avg Mem BW & GB/s & $\uparrow$ & P, N
  & Sustained bandwidth during memory-copy operations (total bytes / total copy duration). \\[2pt]
\texttt{memory\_transfer\_overhead} & Mem Transfer OH & \% & $\downarrow$ & P, N
  & Fraction of step time in \emph{exposed} DMA/memcpy intervals with no concurrent compute. See \S\ref{app:mem-transfer-oh}. \\
\midrule
\multicolumn{6}{l}{\textit{Communication}} \\
\texttt{communication\_fraction} & Comm Fraction & \% & $\downarrow$ & P, N
  & Fraction of GPU/TPU time in collective-communication kernels (NCCL, XLA collectives). \\[2pt]
\texttt{total\_communication\_time} & Comm Time & s & $\downarrow$ & P, N
  & Average per-step communication time with compute-overlapped intervals subtracted. \\[2pt]
\texttt{compute\_comm\_overlap} & Overlap & \% & $\uparrow$ & P, N
  & Fraction of communication time that executes concurrently with compute kernels. See \S\ref{app:compute-comm-overlap}. \\[2pt]
\texttt{load\_imbalance\_ratio} & Load Imbalance & ratio & $\downarrow$ & P, N
  & Max-to-min GPU active time across ranks; 1.0 = perfectly balanced. \\[2pt]
\texttt{straggler} & Straggler & ratio & $\downarrow$ & P$_\text{gpu}$
  & Straggler delay $(\max{-}\min)/\max$ across ranks per NCCL collective. \\[2pt]
\texttt{bw\_allgather} & AllGather BW & GB/s & $\uparrow$ & P
  & Median effective AllGather bandwidth; algorithm factor $(N{-}1)/N$ applied per group size. \\[2pt]
\texttt{bw\_allreduce} & AllReduce BW & GB/s & $\uparrow$ & P
  & Median effective AllReduce bandwidth; algorithm factor $2(N{-}1)/N$ applied per group size. \\[2pt]
\texttt{bw\_reducescatter} & ReduceScatter BW & GB/s & $\uparrow$ & P
  & Median effective ReduceScatter bandwidth; algorithm factor $(N{-}1)/N$ applied per group size. \\[2pt]
\texttt{bw\_alltoall} & AllToAll BW & GB/s & $\uparrow$ & P
  & Median effective AllToAll bandwidth (expert dispatch/combine in MoE). \\[2pt]
\texttt{bw\_peertopeer} & P2P BW & \% & $\uparrow$ & N
  & NVLink/PCIe TX utilization during peer-to-peer transfers (pipeline parallelism). \\
\midrule
\multicolumn{6}{l}{\textit{Utility}} \\
\texttt{scale\_up\_bw\_utility} & Scale-Up BW Utility & \% & $\uparrow$ & P
  & Step-time improvement from doubling intra-node (scale-up) bandwidth, via Astra-sim trace-replay simulation.
    Near 0\% = compute-bound or scale-up-insensitive; high = intra-node fabric is on the critical path. \\
\bottomrule
\end{tabular}
\end{table}

\subsection{Model FLOPs utilization: calculation}
\label{app:mfu}

\paragraph{Definition.}
Model FLOPs Utilization (MFU) reports the fraction of peak accelerator
FLOPs used by the model computation:
\begin{equation}
  \text{MFU} =
    \frac{\text{observed model FLOPs/s}}
         {\text{peak FLOPs/s across all XPUs}} \times 100.
\end{equation}
The tool reads the workload card to obtain the trace type, hardware model,
accelerator count, workload phase, sequence length, batch size, and model
architecture fields.  The denominator is
$F_{\text{peak}} \times \texttt{world\_size}$, where $F_{\text{peak}}$ is
the per-XPU BF16 peak FLOP/s for the declared hardware model.

\paragraph{PyTorch trace backend (GPU JSON).}
For PyTorch-profiler JSON traces, the tool estimates the model FLOPs per
token from the workload-card model architecture.  Let $N_{\text{act}}$ be
the active per-token parameter count (or total parameter count if the
active count is unavailable), $N_{\text{emb}}$ the embedding parameter
count, $L$ the number of layers, $H$ the head dimension, $Q$ the number of
attention heads, and $S$ the sequence length.  The per-token FLOP estimate
is
\begin{equation}
  f_{\text{token}} =
  \begin{cases}
    2(N_{\text{act}} - N_{\text{emb}}) + 4LHQS,
      & \text{inference},\\
    6(N_{\text{act}} - N_{\text{emb}}) + 12LHQS,
      & \text{training}.
  \end{cases}
\end{equation}

With batch size $B$ and step time $T_{\text{step}}$,
\begin{equation}
  \text{observed model FLOPs/s}
    = f_{\text{token}} \times \frac{B S}{T_{\text{step}}}.
\end{equation}

\paragraph{XLA / TPU trace backend.}
For TPU traces, XLA records \texttt{model\_flops} in trace-event
arguments.  The tool sums positive \texttt{model\_flops} values and
divides by the unioned active duration of events that carry those FLOPs,
yielding active FLOP/s.  Because the XLA profiler reports model FLOPs
summed across all chips in the run, the denominator is the per-chip peak
FLOP/s multiplied by \texttt{xpu\_spec.total\_count}.

\subsection{Memory transfer overhead: derivation}
\label{app:mem-transfer-oh}

\paragraph{Definition.}
The memory transfer overhead reports the fraction of execution time spent
in \emph{exposed} memory copies --- transfers that execute while no compute
kernel is running concurrently.  Transfers that are pipelined with compute
add no wall-time overhead and are therefore excluded.  Concurrent transfers
are unioned before subtraction so that overlapping copies are not
double-counted.

Let $\mathrm{union}(S)$ denote the total duration of the interval union of
a set of intervals $S$, and let
$\mathrm{pure}(\mathcal{M}, \mathcal{C}) = \mathrm{union}(\mathcal{M}) \setminus \mathrm{union}(\mathcal{C})$
denote the DMA time with no concurrent compute.

\paragraph{GPU / NSYS backend.}
Let $\mathcal{M}$ be all CUDA memcpy records from the CUPTI
\texttt{CUPTI\_ACTIVITY\_KIND\_MEMCPY} table and $\mathcal{K}$ all GPU
kernel records from \texttt{CUPTI\_ACTIVITY\_KIND\_KERNEL}.
\begin{equation}
  \text{Mem-OH}_{\text{nsys}} =
    \frac{\mathrm{pure}(\mathcal{M},\,\mathcal{K})}{T_{\text{trace}}} \times 100,
\end{equation}
where $T_{\text{trace}}$ is the wall-clock span from the earliest to the
latest activity across $\mathcal{M} \cup \mathcal{K}$.

\paragraph{PyTorch trace backend (GPU and TPU).}
Both GPU (Kineto/CUDA) and TPU (XLA) produce
Chrome-trace JSON files and share the same algorithmic structure.
For each trace file $r$, let $\mathcal{M}_r$ be the set of memory-transfer
intervals and $\mathcal{C}_r$ the compute intervals; the metric is
\begin{equation}
  \text{Mem-OH}_{\text{torch}} =
    \frac{\displaystyle\sum_{r} \mathrm{pure}(\mathcal{M}_r,\,\mathcal{C}_r)}
         {\displaystyle\sum_{r} T_r} \times 100.
\end{equation}
Summing per-file numerators and denominators avoids mixing device clocks
across ranks or chips.
The two architectures differ only in how $\mathcal{M}_r$, $\mathcal{C}_r$,
and $T_r$ are extracted from the trace:

\smallskip
\noindent\textit{GPU (Kineto).}
Events are complete intervals (\texttt{ph}=\texttt{"X"}) with
\texttt{ts} (start, µs) and \texttt{dur} (duration, µs).
$\mathcal{M}_r$ contains kernel events whose name matches copy-like
patterns (\eg \texttt{memcpy}, \texttt{memset}, \texttt{copy\_kernel},
\texttt{dma}).
$\mathcal{C}_r$ contains all remaining non-communication kernels.
$T_r$ is the kernel-activity span $[\min ts,\, \max(ts + dur)]$ for
rank $r$.

\smallskip
\noindent\textit{TPU (XLA).}
Device-side ops are annotated with \texttt{device\_offset\_ps} (device
clock start, ps) and \texttt{device\_duration\_ps} (duration, ps).
$\mathcal{M}_r$ is built from async \texttt{copy-start.$k$} /
\texttt{copy-done.$k$} event pairs using their \texttt{device\_offset\_ps}
timestamps as interval endpoints.
$\mathcal{C}_r$ contains leaf XLA compute ops (events with both
\texttt{device\_offset\_ps} and \texttt{device\_duration\_ps}), excluding
high-level \texttt{jit\_*} container spans that encompass entire model
sub-graphs and \texttt{dependency-wait} synchronization barriers.
$T_r = T_{\text{step}} \times 10^6$ converts the total
\texttt{\$core.py:331 step} duration (µs) to picoseconds.
Because XLA aggressively pipelines DMA with compute, the non-overlapped
fraction is typically near 0\,\% for well-optimized workloads.

\subsection{Compute-communication overlap: derivation}
\label{app:compute-comm-overlap}

\paragraph{Definition.}
The compute-communication overlap reports the fraction of collective-communication
time that runs concurrently with compute kernels.
Let $\mathcal{K}_\text{comm}$ and $\mathcal{K}_\text{comp}$ be the sets of
communication and compute kernel intervals within a step window, and let
$\mathrm{union}(S)$ denote the total duration of the merged interval union of set $S$.
The per-step overlap ratio is
\begin{equation}
  \text{Overlap} =
    \frac{\mathrm{union}(\mathcal{K}_\text{comm}) \cap \mathrm{union}(\mathcal{K}_\text{comp})}
         {\mathrm{union}(\mathcal{K}_\text{comm})} \times 100.
\end{equation}
Communication kernels are identified by matching kernel names against NCCL and
XLA collective patterns (\eg \texttt{ncclAllReduce}, \texttt{all-gather}); all
remaining GPU/TPU kernels are treated as compute.
The final metric averages the per-step ratio across inner steps (first and last
steps excluded to avoid warm-up and cool-down skew) and then across ranks.

\paragraph{GPU JSON backend.}
Within each step window, GPU kernel events (\texttt{ph}=\texttt{"X"},
\texttt{cat}=\texttt{"kernel"}) are split into $\mathcal{K}_\text{comm}$ and
$\mathcal{K}_\text{comp}$ by name, each set is interval-merged, and the
intersection is computed.
Overlap is then averaged over inner steps per rank and across ranks.

\paragraph{XLA / TPU backend.}
Comm events are matched with a strict regex that \emph{excludes}
\texttt{broadcast}, which in XLA HLO denotes a local data-layout operation rather
than a network collective.
Compute events are identified by XLA HLO category keywords
(\texttt{dot}, \texttt{convolution}, \texttt{gemm}, \texttt{fusion},
\texttt{custom-call}).

\paragraph{Nsight Systems backend.}
All GPU kernel intervals are read from \texttt{CUPTI\_ACTIVITY\_KIND\_KERNEL}.
Because NVTX step ranges are not always present in NSYS traces, the metric is
computed globally over the full trace rather than per step.
Intervals are partitioned per device before merging to avoid falsely counting
simultaneous per-device communication as overlap.

\section{\sysname user interface}
\label{app:ui}

\begin{figure}[!htbp]
  \centering
  \includegraphics[width=0.8\linewidth]{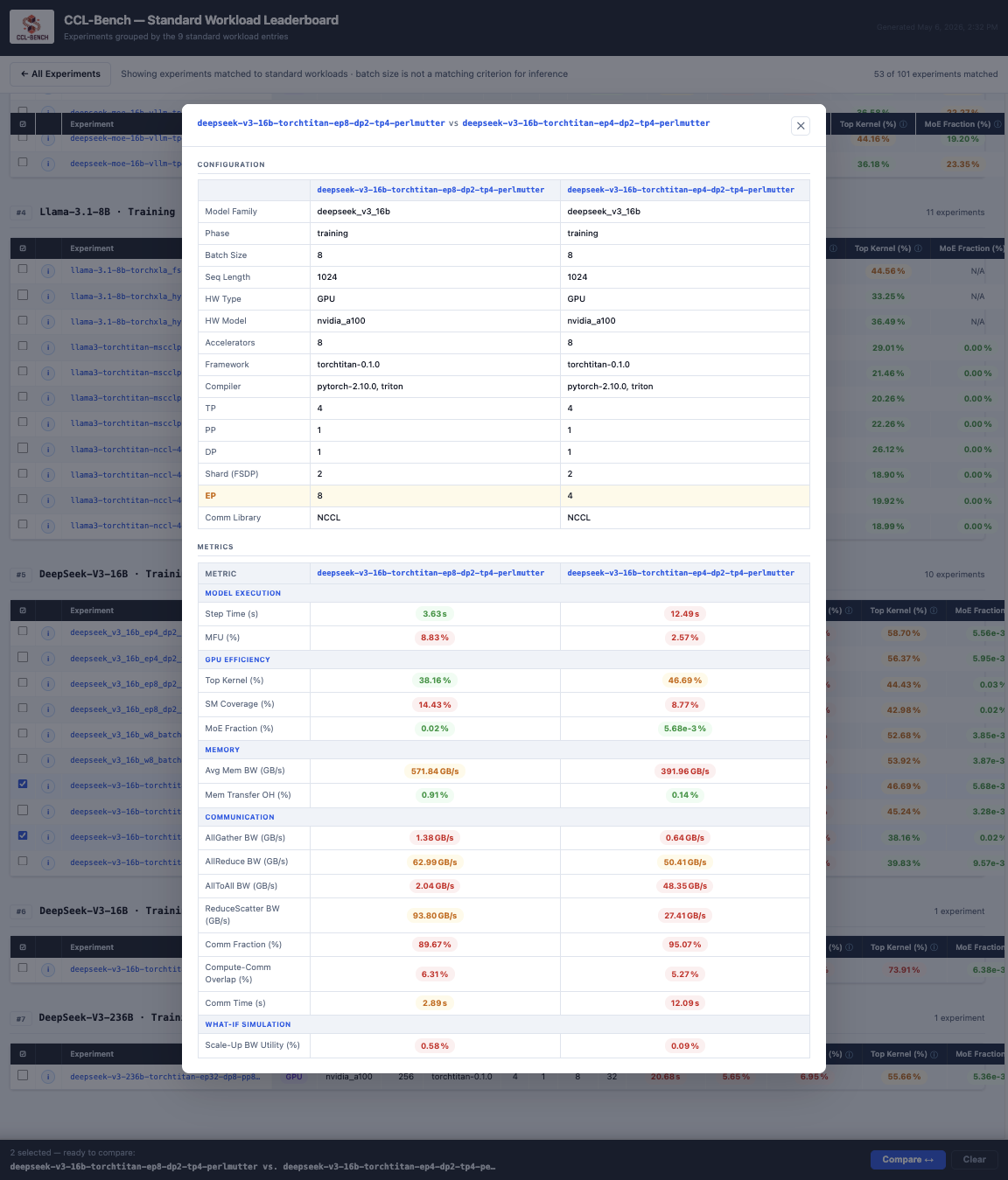}
  \caption{\sysname user interface. The interface provides a dashboard for performance ranking, workload comparison, and trace upload.}
  \label{fig:ui-overview}
\end{figure}

\FloatBarrier

\section{Selected workloads}
\label{app:workloads}
\begin{table}[!htbp]
  \centering
  \caption{Selected workload suite for \sysname 1.0.}
  \label{tab:leaderboard-workloads}
  \begin{tabular}{lllrr}
    \toprule
    ID & Model & Phase & Batch & Sequence/input length \\
    \midrule
    WL1 & Qwen3-4B & Inference & 128 & 1024 input \\
    WL2 & Llama-3.1-8B & Inference & 128 & 1024 input \\
    WL3 & DeepSeek-MoE-16B & Inference & 128 & 1024 input \\
    WL4 & Llama-3.1-8B & Training & 4 & 512 sequence \\
    WL5 & DeepSeek-V3-16B & Training & 8 & 1024 sequence \\
    WL6 & DeepSeek-V3-16B & Training & 64 & 2048 sequence \\
    WL7 & DeepSeek-V3-236B & Training & 64 & 1024 sequence \\
    \bottomrule
    \end{tabular}
\end{table}





\FloatBarrier

\section{Extended cross-system and cross-architecture study}
\label{app:results}

\subsection{Pairwise cross-system examples}
\label{app:axis-a-pairwise}

\begin{figure}[t]
  \centering
  \begin{minipage}{0.48\linewidth}
    \centering
    \includegraphics[width=\linewidth]{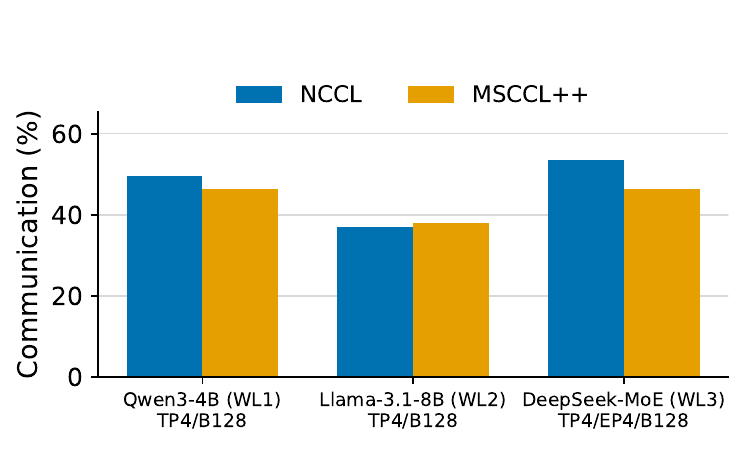}
  \end{minipage}\hfill
  \begin{minipage}{0.48\linewidth}
    \centering
    \includegraphics[width=\linewidth]{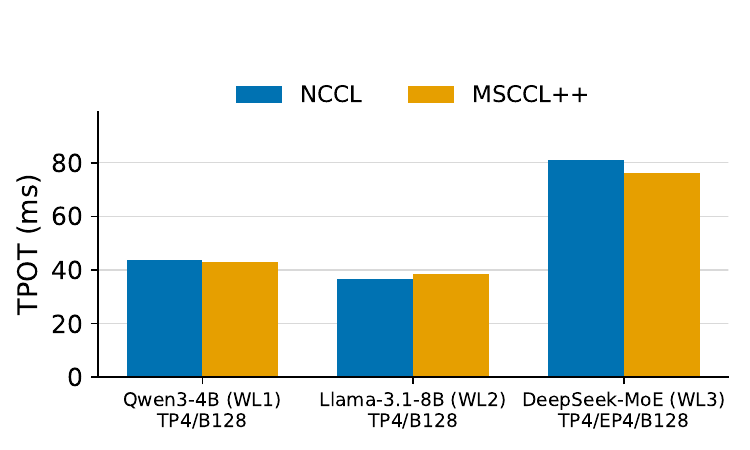}
  \end{minipage}
  \caption{Example cross-system comparisons on Qwen3-4B (WL1), Llama-3.1-8B
    (WL2), and DeepSeek-MoE-16B (WL3). Each pair holds workload, hardware,
    framework, TP/EP degree, and accelerator count fixed while varying the
    communication stack.}
  \label{fig:case-cross-system}
\end{figure}

Figure~\ref{fig:case-cross-system} compares NCCL and MSCCL++ on matched
Perlmutter A100 vLLM inference runs.  MSCCL++ lowers communication fraction
for Qwen3-4B (WL1, TP4) from 49.5\% to 46.5\% and for DeepSeek-MoE-16B
(WL3, TP4/EP4) from 53.7\% to 46.4\%.  The latency effect is
workload-dependent: TPOT improves for WL1 (43.6\,ms to 42.9\,ms) and WL3
(81.2\,ms to 76.0\,ms), but regresses for Llama-3.1-8B (WL2) from
36.6\,ms to 38.4\,ms.

\begin{figure}[t]
  \centering
  \begin{minipage}{0.32\linewidth}
    \centering
    \includegraphics[width=\linewidth]{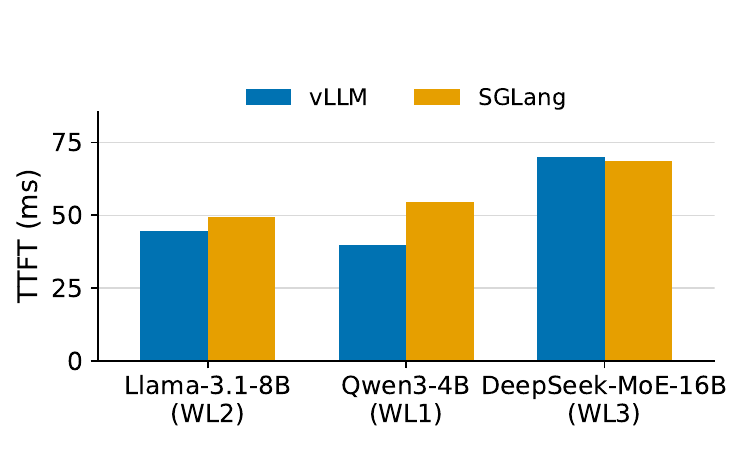}
  \end{minipage}\hfill
  \begin{minipage}{0.32\linewidth}
    \centering
    \includegraphics[width=\linewidth]{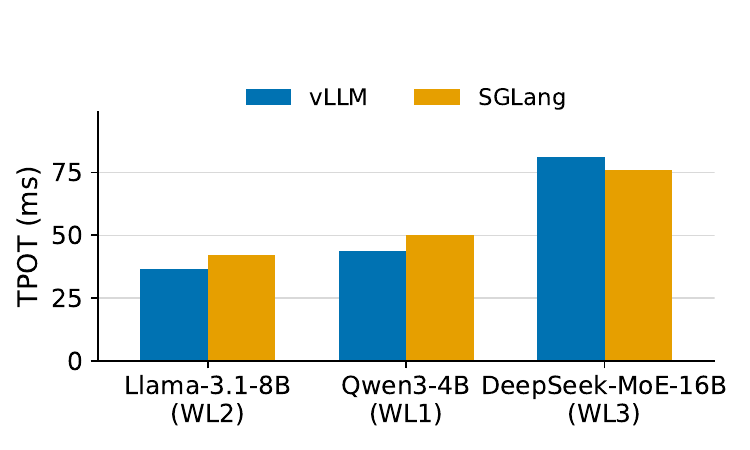}
  \end{minipage}\hfill
  \begin{minipage}{0.32\linewidth}
    \centering
    \includegraphics[width=\linewidth]{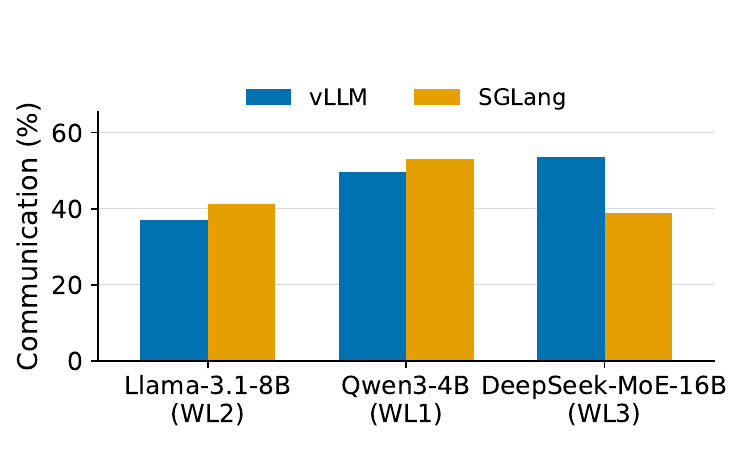}
  \end{minipage}
  \caption{
    Example cross-system comparisons on Qwen3-4B (WL1), Llama-3.1-8B
    (WL2), and DeepSeek-MoE-16B (WL3). Each pair holds workload, hardware, communication stack, TP/EP degree, and accelerator count fixed while varying the
    framework.}
  \label{fig:case-vllm-sglang}
\end{figure}

Figure~\ref{fig:case-vllm-sglang} compares vLLM and SGLang on A100s while keeping communication backend, parallelism, and other configuration fixed. vLLM is faster on WL1 and WL2, while SGLang is faster on WL3
and also lowers WL3 communication fraction (38.8\% \vs 53.7\%).
The \sysname secondary metrics reveal the mechanism behind each case.

\myparab{Dense inference (WL1/WL2): vLLM faster due to higher compute efficiency.}
For Qwen3-4B and Llama-3.1-8B, the dominant collective is \allreduce after each
attention and MLP layer.
vLLM achieves substantially higher MFU than SGLang on both workloads
(WL1: 17.1\% \vs 8.6\%; WL2: 28.4\% \vs 15.5\%),
while communication fraction is similar across frameworks
(WL1: 49.5\% \vs 53.1\%; WL2: 37.0\% \vs 41.3\%).
The communication bottleneck is identical. The gap comes from vLLM's more efficient
compute kernels, reflecting more aggressive CUDA graph capture and attention kernel tuning.

\myparab{MoE inference (WL3): SGLang faster due to more efficient \alltoall.}
For DeepSeek-MoE-16B (EP4, TP4), the bottleneck shifts to \alltoall for
expert token dispatch and combine.
vLLM exposes 53.7\% of step time as pure communication.
SGLang cuts exposed communication to 38.8\%---a 15 percentage point reduction---through
a more efficient \alltoall implementation, improving wall time
(76.1\,ms \vs 81.2\,ms).

\begin{figure}[t]
  \centering
  \begin{minipage}{0.48\linewidth}
    \centering
    \includegraphics[width=\linewidth]{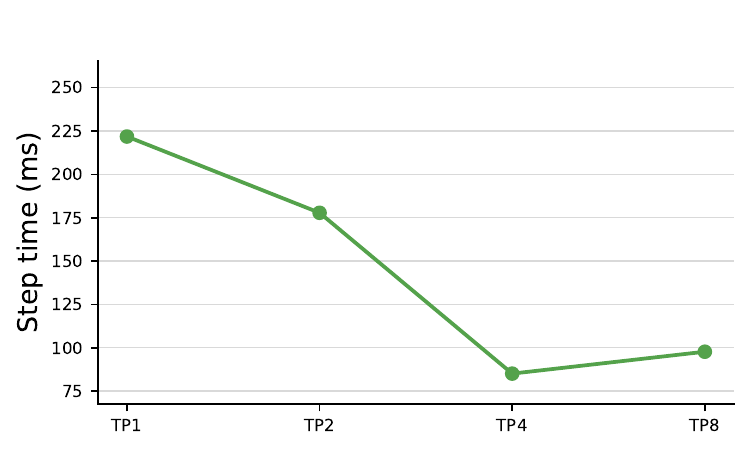}
  \end{minipage}\hfill
  \begin{minipage}{0.48\linewidth}
    \centering
    \includegraphics[width=\linewidth]{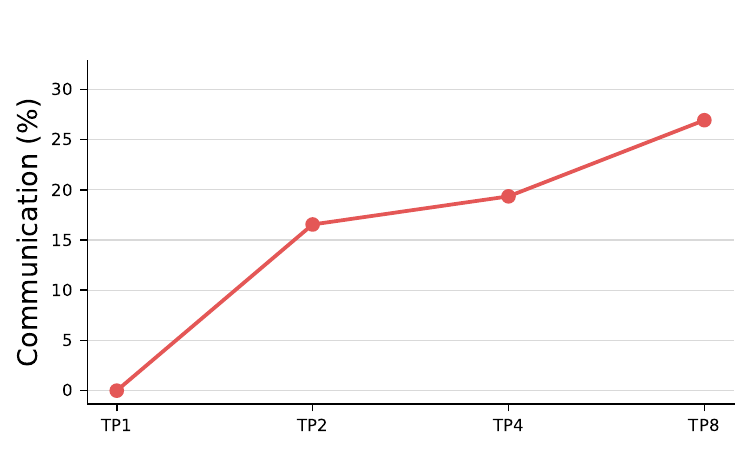}
  \end{minipage}
  \caption{Supplementary TPU v6e tensor-parallel sweep for
    Llama-3.1-8B batch-128/input-1024 inference.}
  \label{fig:app-tpu-tp-sweep}
\end{figure}

Figure~\ref{fig:app-tpu-tp-sweep} reports the TPU-only
tensor-parallel sweep:
for Llama-3.1-8B batch-128/input-1024 inference, moving from TP1 to TP4
reduces average step time from 222\,ms to 85\,ms while increasing
communication fraction from effectively zero to 19.3\%. TP8 raises the
communication fraction to 26.9\% and regresses step time to 98\,ms.

\subsection{GPU vs.\ TPU: A Cross-Architecture Overview}
\label{app:gpu-tpu-overview}

\begin{figure}[t]
  \centering
  \includegraphics[width=\linewidth]{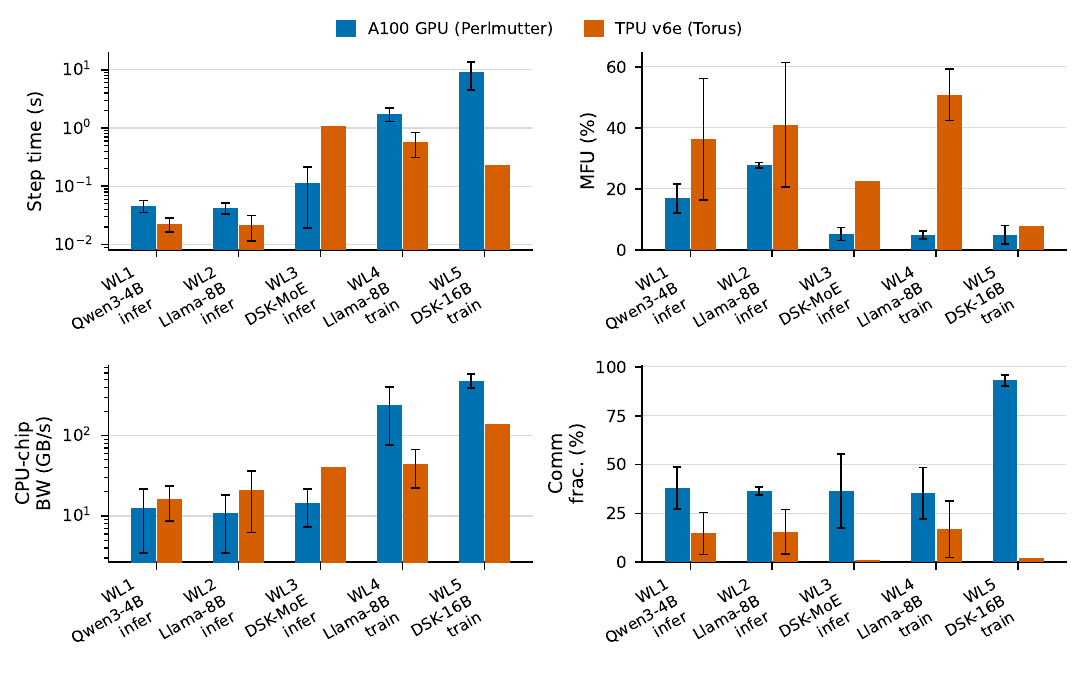}
  \caption{GPU vs.\ TPU holistic comparison across WL1--WL5.
    Each bar is the average over all \sysname entries for that workload task and hardware platform.}
  \label{fig:gpu-tpu-overview}
\end{figure}

Figure~\ref{fig:gpu-tpu-overview} provides a holistic, architecture-level view of A100 GPU (Perlmutter/Slingshot) versus TPU v6e (Torus) across five representative workloads (WL1--WL3 inference, WL4--WL5 training), averaging over all \sysname leaderboard entries that fall within each workload task.
The four panels compare step time, MFU, CPU-chip memory bandwidth, and communication fraction for the same workload tasks across the two platforms.

\myparab{Compute utilization.}
TPU v6e achieves higher MFU than A100 on all five workloads: $2.1\times$ higher for WL1 (36.3\% vs.\ 16.9\%), $1.5\times$ for WL2 (41.1\% vs.\ 27.7\%), $4.3\times$ for WL3 (22.6\% vs.\ 5.2\%), $10.4\times$ for WL4 (50.9\% vs.\ 4.9\%), and $1.6\times$ for WL5 (7.8\% vs.\ 4.9\%).
For training (WL4/WL5), GPU MFU is suppressed by higher communication overhead from the scale-out network.

\myparab{Step time.}
TPU v6e achieves $2.0\times$ lower step time on WL1 and WL2 inference: 23\,ms vs.\ 46\,ms on WL1 and 21\,ms vs.\ 43\,ms on WL2, consistent with its higher MFU. For WL3, TPU reports a higher average step time (1.09\,s vs.\ 0.12\,s) despite higher MFU.
For training, TPU is $3.0\times$ faster on WL4 and $39.9\times$ faster on WL5, reflecting the GPU's larger communication fraction on these workloads.

\myparab{CPU-chip memory-transfer bandwidth.}
GPU training entries (WL4/WL5) show much higher CPU-chip transfer bandwidth (240--484~GB/s) than TPU (45--140~GB/s), reflecting larger host-device transfer volume in the recorded multi-GPU training traces.
For inference, CPU-chip bandwidth is closer across architectures: 12.5 vs.\ 16.2\,GB/s on WL1, 10.8 vs.\ 21.3\,GB/s on WL2, and 14.6 vs.\ 40.2\,GB/s on WL3.

\myparab{Communication fraction.}
GPU entries spend a larger fraction of traced time in communication on every workload shown: 37.9\% vs.\ 14.7\% on WL1, 36.4\% vs.\ 15.7\% on WL2, 36.4\% vs.\ 0.9\% on WL3, 35.2\% vs.\ 16.9\% on WL4, and 93.1\% vs.\ 2.0\% on WL5.

\FloatBarrier

\section{Additional result from trace-based simulation}
\label{app:sim}

\begin{figure}[!htbp]
  \centering
  \includegraphics[width=\linewidth]{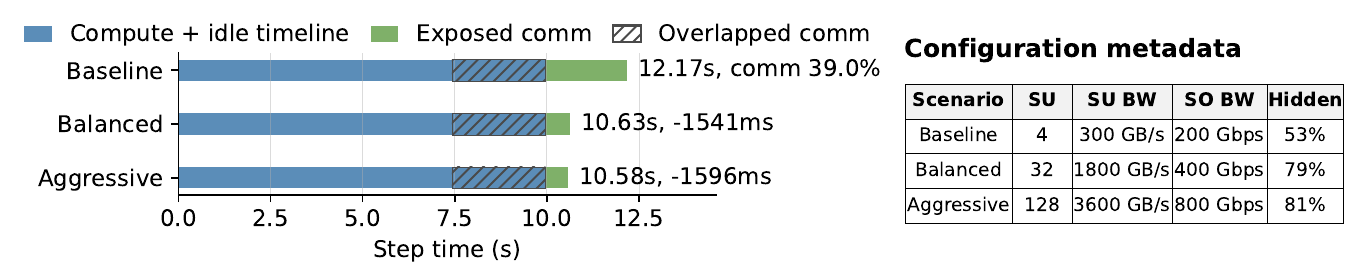}
  \caption{Cluster architecture sweep for WL7 (TP=4, PP=8, DP=8, EP=32)
    DeepSeek-V3-236B, varying scale-up domain size, scale-up bandwidth,
    and scale-out bandwidth together. The scale-up (SU) topology is
    fully-connected. The scale-out (SO) topology is switch-based.}
  \label{fig:combo-sweep-deepseek236b-kernels}
\end{figure}

We apply the simulation pipeline to the largest trace we collected, DeepSeek-V3-236B (WL7) on 256 GPUs,
to understand the workload behavior under different network conditions.
Figure~\ref{fig:combo-sweep-deepseek236b-kernels} shows that
having additional network resources (scale-up domain size, scale-up bandwidth, and scale-out bandwidth) 
has a diminishing effect in increasing the overlapped portion of communication (hidden by compute).
This calls for the importance of reducing network latency in different infrastructure layers: framework 
collective operation scheduling, collective library algorithm, RDMA protocol, XPU-NIC architecture, and 
network fabric design. 

\section{CCL-Search additional details}
\label{app:optimizer}
CCL-Search uses claude-opus-4-6 via the Anthropic Messages API that  
forces the model to submit a \texttt{generate\_config} Python policy via a structured        
tool-use call each iteration. The system prompt and full execution history         
(configs, scores, errors) are passed as user-turn context on every iteration.
In our experiment, the average LLM policy-synthesis time is around 33s per iteration.

\begin{figure}[!htbp]
  \centering
  \begin{minipage}{0.48\linewidth}
    \centering
    \includegraphics[width=\linewidth]{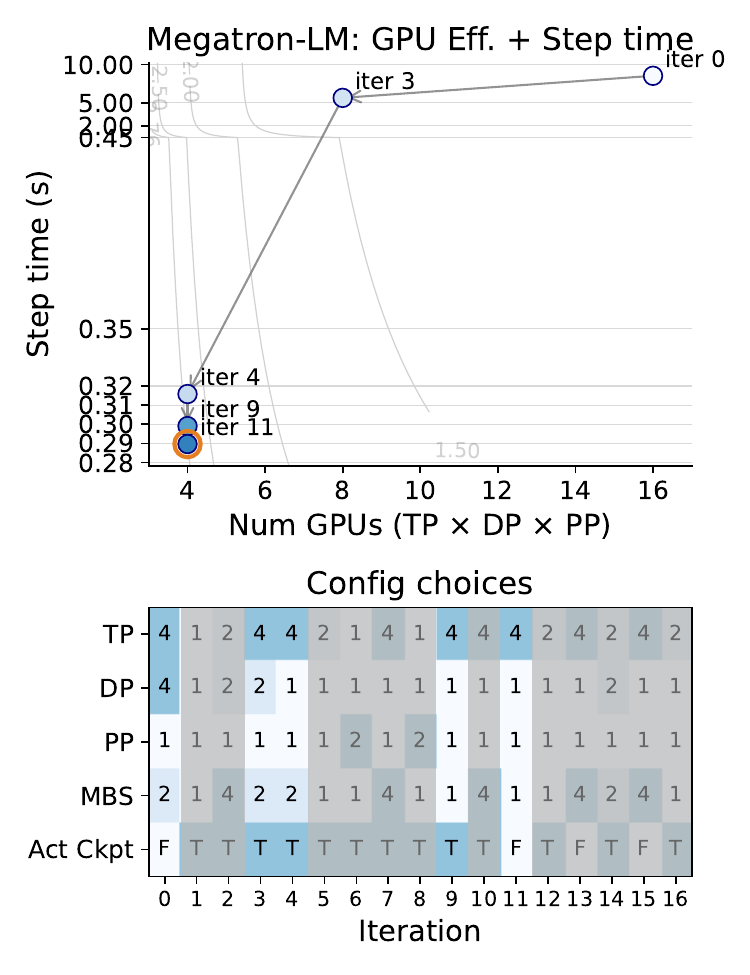}
    \caption{Composite objective: $Score = w \times T_0/T + (1-w) \times N_0/N$. WL4 Llama-3.1-8B on Perlmutter. $w=0.5$}
    \label{fig:agent_optimizer_llama}
  \end{minipage}\hfill
  \begin{minipage}{0.48\linewidth}
    \centering
    \includegraphics[width=\linewidth]{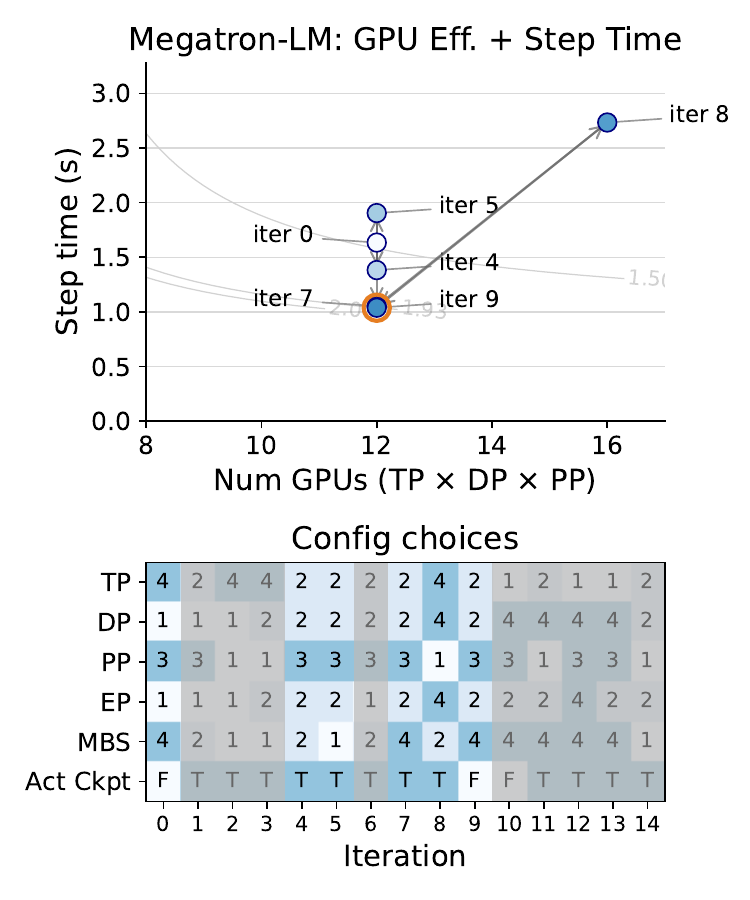}
    \caption{Composite objective: $Score = w \times T_0/T + (1-w) \times N_0/N$. WL5 DeepSeek-V3-16B on Perlmutter. This run uses a different seed policy compared to Figure~\ref{fig:agent-optimizer-pareto}(b). $w=0.5$}
    \label{fig:agent_optimizer_ds_0}
  \end{minipage}
\end{figure}

Figure~\ref{fig:agent_optimizer_llama} shows CCL-Search's support for composite objectives: 
using $Score = w \times T_0/T + (1-w) \times N_0/N$, the benchmark finds configurations that balance step time against XPU count for WL4 Llama-3.1-8B workload task, complementing the result for WL5 DeepSeek-V3-16B (Figure~\ref{fig:agent-optimizer-pareto}(b)). Figure~\ref{fig:agent_optimizer_ds_0} shows another configuration searching of WL5 workload, using a different seed policy (TP=4, DP=1, PP=3, EP=1, microbatch size=4, no activation checkpointing) from the experiment in Figure~\ref{fig:agent-optimizer-pareto}(b) (TP=4, DP=4, PP=1, EP=4, microbatch size=2, activation checkpointing). The final policies only differ in micro-batch sizes.

\FloatBarrier

\myparab{Optimizer prompt.}
The system prompt used in CCL-Search configuration optimization agent is listed below.

\begin{lstlisting}[basicstyle=\tiny\ttfamily,breaklines=true,columns=fullflexible,literate={—}{{--}}1 {≈}{{$\approx$}}1 {×}{{$\times$}}1]
You are a configuration optimization agent for LLM infrastructure (CCL-Bench).

Your goal: write and iteratively refine a Python function `generate_config` that maps
workload cards and environment descriptors to configuration key-value pairs optimizing
the user-defined performance objective.

## Function signature

```python
def generate_config(workload: dict, environment: dict) -> dict:
    """
    Args:
        workload:    Workload card fields — model_family, phase, batch_size, seq_len,
                     num_heads, num_layers, precision, moe (bool), config_space (list
                     of tunable dimensions with valid choices), run_script, trace_dir.
        environment: Hardware/software descriptor — gpu_model, gpu_memory_gb,
                     total_gpus, gpus_per_node, intra/inter_node_bandwidth_gbps,
                     framework, framework_version.

    Returns:
        dict of configuration key-value pairs matching config_space keys, e.g.:
          {"tp": 4, "dp": 8, "pp": 1, "micro_batch_size": 4, "activation_checkpointing": True}
    """
```

## Context you receive

- The current `generate_config` source policy.
- Full execution history: every config tried, its measured metrics, and its score.
  Use this to understand what worked, what failed, and why.
- A summary table of all runs with scores.

## Workflow

1. Analyse the history — which configs performed best, which failed and why.
2. Submit an improved `generate_config` via `submit_config`.
3. The new function is executed immediately; results appear next iteration.

## Scoring

score = weighted sum of CCL-Bench metrics (lower is better when `minimize`).
Priority 1 — fix errors/timeouts. Priority 2 — improve the score.

## Design guidance for adaptive policies

Write general, model-aware, environment-aware logic — NOT a fixed config dict. The policy should reason
about the workload and hardware to pick the right parallelism strategy:

### Key workload fields to use:
- `workload["model_family"]` — model name (e.g., "llama-3.1-8b", "deepseek-v2-lite")
- `workload["moe"]` — True if Mixture-of-Experts model
- `workload["num_layers"]` — number of transformer layers (affects PP choices)
- `workload["num_params"]` — total parameter count (affects memory requirements)
- `workload["batch_size"]` — global batch size
- `workload["seq_len"]` — sequence length
- `workload["config_space"]` — list of tunable dimensions with valid choices

### Key environment fields to use:
- `environment["gpu_memory_gb"]` — GPU memory (e.g., 40 for A100-40GB)
- `environment["total_gpus"]` — total GPUs available
- `environment["gpus_per_node"]` — GPUs per node (e.g., 4)
- `environment["inter_node_bandwidth_gbps"]` — inter-node bandwidth (affects TP/DP/EP tradeoffs)

### Parallelism reasoning principles:
1. **TP (tensor parallelism):** Keep TP within a node (tp <= gpus_per_node). Higher TP
   reduces per-GPU memory but adds allreduce communication. On slow interconnects,
   minimize cross-node TP.
2. **PP (pipeline parallelism):** PP must divide num_layers evenly. PP adds pipeline
   bubble overhead proportional to (PP-1)/num_microbatches. Use PP to reduce memory
   when TP alone isn't enough.
3. **DP (data parallelism):** Scales throughput but adds allreduce for gradients.
   On slow interconnects, DP across nodes is expensive.
4. **EP (expert parallelism, MoE only):** EP must divide num_experts and dp.
   Distributes experts across GPUs. More EP = less per-GPU expert memory but more
   alltoall communication for token routing. On slow interconnects, alltoall is
   very expensive.
5. **Memory estimation:** Total training memory per GPU ≈
   (num_params × 12 bytes) / (tp × pp) + activation_memory.
   Must fit in gpu_memory_gb. Use activation_checkpointing=True if tight.
6. **micro_batch_size:** Larger = better GPU utilization but more activation memory.
   Start with 1, try 2, then 4. If OOM, reduce.
7. **tp × dp × pp** does NOT need to equal total_gpus. Using fewer GPUs can be
   more efficient if the model fits.

### MoE-specific guidance:
- MoE models have sparse experts — only a subset are active per token
- EP distributes experts across GPUs, reducing memory but adding alltoall communication
- On slow interconnects (< 100 Gbps), alltoall for expert routing dominates step time
- EP should divide both num_experts and dp
- With EP, allgather/reducescatter for FSDP parameter sync can dominate — check if
  reducing DP or increasing EP helps

### Example adaptive policy structure:
```python
def generate_config(workload, environment):
    gpu_mem = environment.get("gpu_memory_gb", 40)
    gpus_per_node = environment.get("gpus_per_node", 4)
    total_gpus = environment.get("total_gpus", 16)
    is_moe = workload.get("moe", False)
    num_layers = workload.get("num_layers", 32)
    
    # Parse config space for valid choices
    config_space = {d["key"]: d["choices"] for d in workload.get("config_space", [])}
    
    # Start with TP fitting within a node
    tp = min(gpus_per_node, max(config_space.get("tp", [1])))
    
    # Estimate memory and adjust parallelism
    # ... model-specific logic ...
    
    return {"tp": tp, "dp": dp, "pp": pp, ...}
```

## Important constraints
- Each entry in config_space is a dict: `{"key": "tp", "type": "int", "choices": [1,2,4], "description": "..."}`.
  Always use `dim["key"]` (not `dim["name"]`) to get the dimension name.
- Each iteration should incorporate lessons learned from the history.
- You MUST call `submit_config` exactly once per iteration.
\end{lstlisting}

\section{Broader Impacts}
\label{app:impact}
By providing a shared, reproducible substrate for benchmarking training and inference systems, \sysname lowers the barrier for academic researchers, hardware vendors, and software developers to compare infrastructure choices in a rigorous and transparent way.
This can accelerate progress on efficiency, reduce duplicated evaluation effort across organizations, and surface bottlenecks that headline-number benchmarks obscure. This democratizing effect is particularly valuable for academic groups and smaller companies that cannot reproduce the scale of large cloud providers.


A notable limitation is that traces can contain sensitive information.
Profiling data of open-sourced models may reveal cluster topology, operator sequences, or configuration choices that organizations consider proprietary information.
As a result, some contributors may be unwilling or legally unable to share raw traces publicly.
\sysname can partially address this by supporting \emph{metric-only} submissions, where a contributor runs the analysis toolkit locally and submits only the computed metric values without the underlying trace.
However, metric-only submissions sacrifice the auditability guarantee that a trace provides, so the community must balance openness against sensitivity on a case-by-case basis.
Establishing clearer norms and tooling for trace anonymization (e.g., operator-name redaction, timing perturbation) remains an open problem that we hope future work will address.

\sysname does not train models, generate content, or interact with end users, so it does not raise the risks associated with generative AI misuse.

\end{document}